\documentclass[dvips]{acta}
\usepackage{supertabular,lscape,epsfig}
\usepackage{amssymb}
\usepackage{amsmath}

\SetPages{0}{0}

\SetVol{00}{2014}
\begin{document}

\renewcommand*{\thefootnote}{\fnsymbol{footnote}}
\setcounter{footnote}{1}
\begin{Titlepage}
\Title{A CCD Search for Variable Stars of Spectral Type B\\ in the Northern Hemisphere Open Clusters.\\ IX. NGC 457\footnote{Based on observations obtained with the Apache Point Observatory 3.5-meter telescope, which is owned and operated by the Astrophysical Research Consortium, and with Bia{\l}k\'ow and Ostrowik 60-cm reflecting telescopes operated by University of Wroc{\l}aw and University of Warsaw, respectively.}
}

\Author{D. ~ M~o~\'z~d~z~i~e~r~s~k~i, ~ A. ~ P~i~g~u~l~s~k~i, ~ G. ~ K~o~p~a~c~k~i,\\ Z.~~K~o~{\l}~a~c~z~k~o~w~s~k~i ~ and ~ M.~~S~t~\k{e}~\'s~l~i~c~k~i}
{Astronomical Institute, University of Wroc{\l}aw, Kopernika 11, 51-622 Wroc{\l}aw\\
e-mail: mozdzierski@astro.uni.wroc.pl}

\end{Titlepage}
\setcounter{footnote}{0}
\renewcommand*{\thefootnote}{\arabic{footnote}}

\Abstract{We present results of a $BVI_{\rm C}$ variability survey in the young open cluster NGC\,457 based on observations obtained during three separate runs spanning almost 20 years. In total, we found 79 variable stars, of which 66 are new. The $BVI_{\rm C}$ photometry was transformed to the standard system and used to derive cluster parameters by means of isochrone fitting. The cluster is about 20~Myr old, the mean reddening amounts to about 0.48~mag in terms of the color excess $E(B-V)$. Depending on the metallicity, the isochrone fitting yields a distance between 2.3 and 2.9~kpc, which locates the cluster in the Perseus arm of the Galaxy. 

Using the complementary H$\alpha$ photometry carried out in two seasons separated by over 10 years, we find that the cluster is very rich in Be stars. In total, 15 stars in the observed field of which 14 are cluster members showed H$\alpha$ in emission either during our observations or in the past. Most of the Be stars vary in brightness on different time scales including short-period variability related most likely to $g$-mode pulsations. A single-epoch spectrum of NGC\,457-6 shows that this Be star is presently in the shell phase.

The inventory of variable stars in the observed field consists of a single $\beta$~Cep-type star, NGC\,457-8, 13 Be stars, 21 slowly pulsating B stars, seven $\delta$~Sct stars, one $\gamma$~Dor star, 16 unclassified periodic stars, 8 eclipsing systems and a dozen of stars with irregular variability, of which six are also B-type stars. As many as 45 variable stars are of spectral type B which is the largest number in all open clusters presented in this series of papers. The most interesting is the discovery of a large group of slowly pulsating B stars which occupy the cluster main sequence in the range between $V=$ 11 and 14.5~mag, corresponding to spectral types B3 to B8. They all have very low amplitudes and about half show pulsations with frequencies higher than 3~d$^{-1}$. We argue that these are most likely fast-rotating slowly pulsating B stars, observed also in other open clusters.}
{open clusters: individual: NGC 457 -- stars: $\beta$~Cep -- stars: SPB -- stars: $\delta$~Sct}

\section{Introduction}
Since 1994 we have been running a program of searching for variable stars of spectral type B in open clusters (Jerzykiewicz {\etal}2011 and references therein). The most important goal of the program is the characterization of young open clusters in terms of the incidence of the hot pulsating stars and B stars with emission in H$\alpha$ (Be stars) and its dependence on cluster parameters (age, metallicity). The results of our search can be used, for instance, for mapping the observational instability strips of $\beta$~Cep and slowly pulsating B (SPB) stars. In addition, we are looking for clusters which are suitable for ensemble asteroseismology (see Saesen {\etal}2010b), i.e.~are rich in pulsating stars and containing bright eclipsing binaries. The latter can be used to derive precise stellar parameters of some member stars and put additional constraints on all members, especially pulsating stars. Our program already resulted in finding that the open cluster NGC\,6910 is a very good object for this purpose (
Ko{\l}aczkowski {\etal}2004, Saesen {\etal}2010b).

In the present paper, we describe the results of the variability survey in the young open cluster NGC~457 (C\,0115+580, OCl 321). The cluster is located in the Perseus arm of the Galaxy ($\alpha_\textrm{2000}=\mbox{1}^{\rm h}\mbox{19.7}^{\rm m}$, $\delta_\textrm{2000}=+$58$^\circ$17$^\prime$, $l=$ 126.6$^\circ$, $b=-$4.4$^\circ$) and belongs to the Cassiopeia-Perseus complex of open clusters (Trumpler 1930, Alter 1944b, de la Fuente Marcos and de la Fuente Marcos 2009). The field of the cluster is dominated by two bright supergiants, $\phi$~Cas (F0\,Ia, $V=$ 5.0~mag) and HD\,7902 (B6\,Ib, $V=$ 7.0~mag) which form `eyes' responsible for the Owl and ET nicknames of the cluster. There is also a red supergiant in the field of the cluster, HDE\,236697 $=$ V466~Cas (M1.5\,Iab, $V=$ 8.7~mag).  The radial velocities and spectroscopic parallaxes of all three supergiants are consistent with the cluster membership (Harris 1976, Eggen 1982, Sowell 1987, Liu {\etal}1991). 

NGC 457 has been frequently studied photometrically. The photographic studies include two-band International System photometry of Alter (1944a) and Bod\'en (1946, 1950, 1951), $RGU$ photometry of Becker and Stock (1954) and $UBV$ photometry of Pesch (1959), Hoag {\etal}(1961), Jones and Hoag (1968) and Moffat (1972). The $UBV$ photoelectric photometry of some stars in the cluster was carried out by Hoag {\etal}(1961) and Jones and Hoag (1968) while Eggen (1982) made the first Str\"omgren photometry of seven cluster members. Finally, CCD photometry of NGC\,457 was provided by Fitzsimmons (1993) in Str\"omgren and Phelps and Janes (1994) and Zhang {\etal}(2012) in $UBV$ systems, respectively.

The best distance estimates for NGC\,457 range  between 2.4 and 2.9~kpc (Pesch 1959, Lindoff 1968, Mermilliod 1981, Kharchenko {\etal}2005, Maciejewski {\etal}2008) locating the cluster in the Perseus arm, as mentioned above. The age of the cluster amounts to about 20 Myr (Meynet {\etal}1993, Phelps and Janes 1994, Loktin {\etal}2001, Tadross 2001, Kharchenko {\etal}2005, Maciejewski {\etal}2008, Zhang {\etal}2012). The cluster is moderately reddened with $E(B-V) \approx$ 0.5~mag (Pesch 1959, Johnson {\etal}1961, Baade 1983, Meynet {\etal}1993, Phelps and Janes 1994, Loktin {\etal}2001, Kharchenko {\etal}2005, Maciejewski {\etal}2008) and the reddening is almost constant across the cluster.

The paper is organized as follows. In Section 2 we describe our photometric observations of NGC\,457 and the methods of reduction. Then (Sect.~3) we give an account of standardization of the $BVI_{\rm C}$ photometry and derive global parameters of the cluster. The H$\alpha$ photometry and the variability of emission in H$\alpha$ is summarized in Sect.~4, while in Sect.~5 supplementary spectroscopy of three cluster members obtained in the Apache Point Observatory is presented. Finally, we make a review of variable stars in the cluster (Sect.~6) and finish with a discussion and conclusions (Sect.~7).

\begin{figure}[ht]
\includegraphics[width=\textwidth]{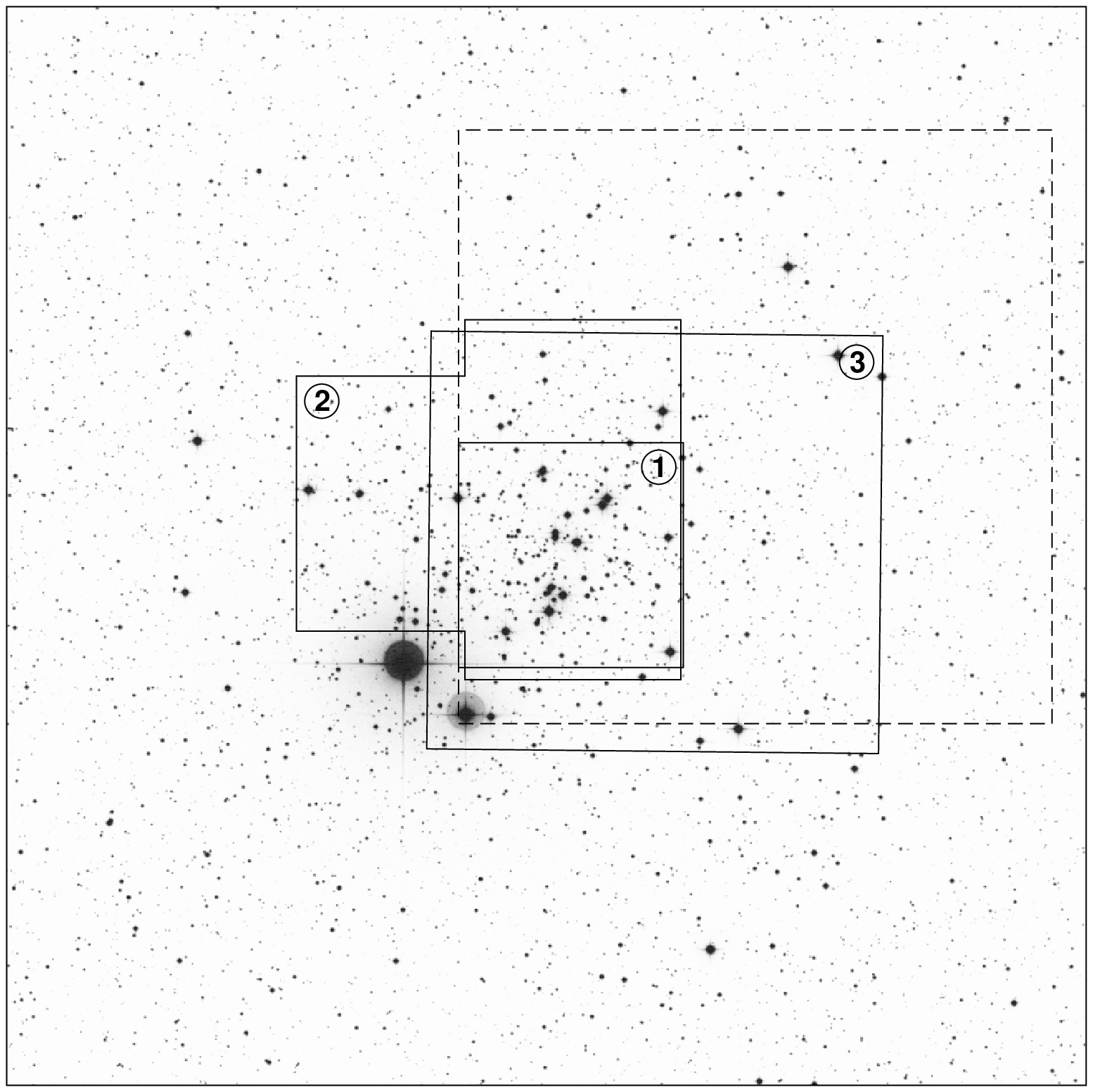}
\FigCap{A 30$^\prime\times$30$^\prime$ fragment of the POSS-II blue plate centered on NGC\,457. Solid lines delimit the fields observed during the three runs (labeled), dashed, the field covered by Zhang {\etal}(2012); see text for explanation. North is up, east to the left.}
\label{poss}
\end{figure}

\section{Time-Series Photometry}
Photometric time-series observations of NGC\,457 were obtained during three runs. The first run consisted of only four nights in 1993 in the Ostrowik station, University of Warsaw. The observations were carried out by one of us (GK) with a 60-cm reflecting telescope and the attached CCD camera covering a 6.3$^\prime\times$ 6.3$^\prime$ field of view (Udalski and Pych 1992, the field labeled with encircled `1' in Fig.\,1). These observations were followed by the observations made in the years 1999\,--\,2002 in the Bia{\l}k\'ow station, University of Wroc{\l}aw, during 31 observing nights with a CCD camera attached to a similar 60-cm reflecting telescope. A single field of view covered 6$^\prime\times$ 4$^\prime$ and for this reason five overlapping fields were observed (the field labeled with encircled `2' in Fig.\,1). During this run, NGC\,457 was usually observed as a secondary target, and therefore the observations rarely spanned a significant part of a night. Finally, the third run was carried out also in 
Bia{\l}k\'ow observatory during 24 nights between December 4, 2010 and March 12, 2011. We used the same telescope and Andor Tech.~DW432-BV back-illuminated CCD camera covering 13$^\prime\times$ 12$^\prime$ field of view (labeled with encircled `3' in Fig.\,1). Over 4500 CCD frames through $B$, $V$, and $I_\textrm{C}$ filters of the Johnson-Kron-Cousins $UBV(RI)_\textrm{C}$ photometric system were taken during this run. The exposure times ranged from 30 to 180 s, depending on filter and the sky conditions. The fields covered during the three runs are shown in Fig.~1 and labeled sequentially. As can be seen from the figure, they partially overlap. In all runs, we avoided the very bright star $\phi$~Cas.

All observations were calibrated in a standard way, which included bias and dark frame subtraction as well as flat-fielding. For each frame, we calculated aperture and profile magnitudes of the stars using the {\sc Daophot II} package (Stetson 1987), and then derived differential magnitudes which were subject of subsequent time-series analysis. In total, we identified 2940 stars in the fields observed during all three runs. Although for NGC\,457 numerous numbering systems exist, throughout the paper we will consistently use the numbering system provided by the WEBDA\footnote{The WEBDA database is available at \textit{http://www.univie.ac.at/webda/webda.html}.} database.

\section{$BVI_\textrm{\textup{C}}$ Photometry and the Parameters of NGC\,457}
In order to be able to derive cluster parameters, we transformed our average instrumental magnitudes and color indices to the standard system. This was done using photometry of Phelps and Janes (1994) as a standard. The following transformation equations were obtained:
\begin{equation}
 V=v-(0.058\pm 0.004)\times(v-i)+(13.488\pm 0.002), \quad \sigma=0.026~\mbox{mag},
\end{equation}
\begin{equation}
(B-V)=(1.262\pm 0.008)\times(b-v)+(0.388\pm 0.002), \quad \sigma=0.027~\mbox{mag},
\end{equation}
\begin{equation}
(V-I_{\textup{C}})=(0.998\pm 0.006)\times(v-i)+(0.506\pm 0.003), \quad \sigma=0.032~\mbox{mag},
\end{equation}
where the uppercase letters denote standard magnitudes, lowercase ones, instrumental magnitudes and $\sigma$ is the standard deviation of the fit. The coefficients of transformation equations were obtained by means of the least squares method using 360, 252 and 222 stars for the above three equations, respectively. The resulting color-magnitude (CM) diagrams are shown in Fig.\,2.

\begin{figure}[ht]
\includegraphics[angle=270,width=\textwidth]{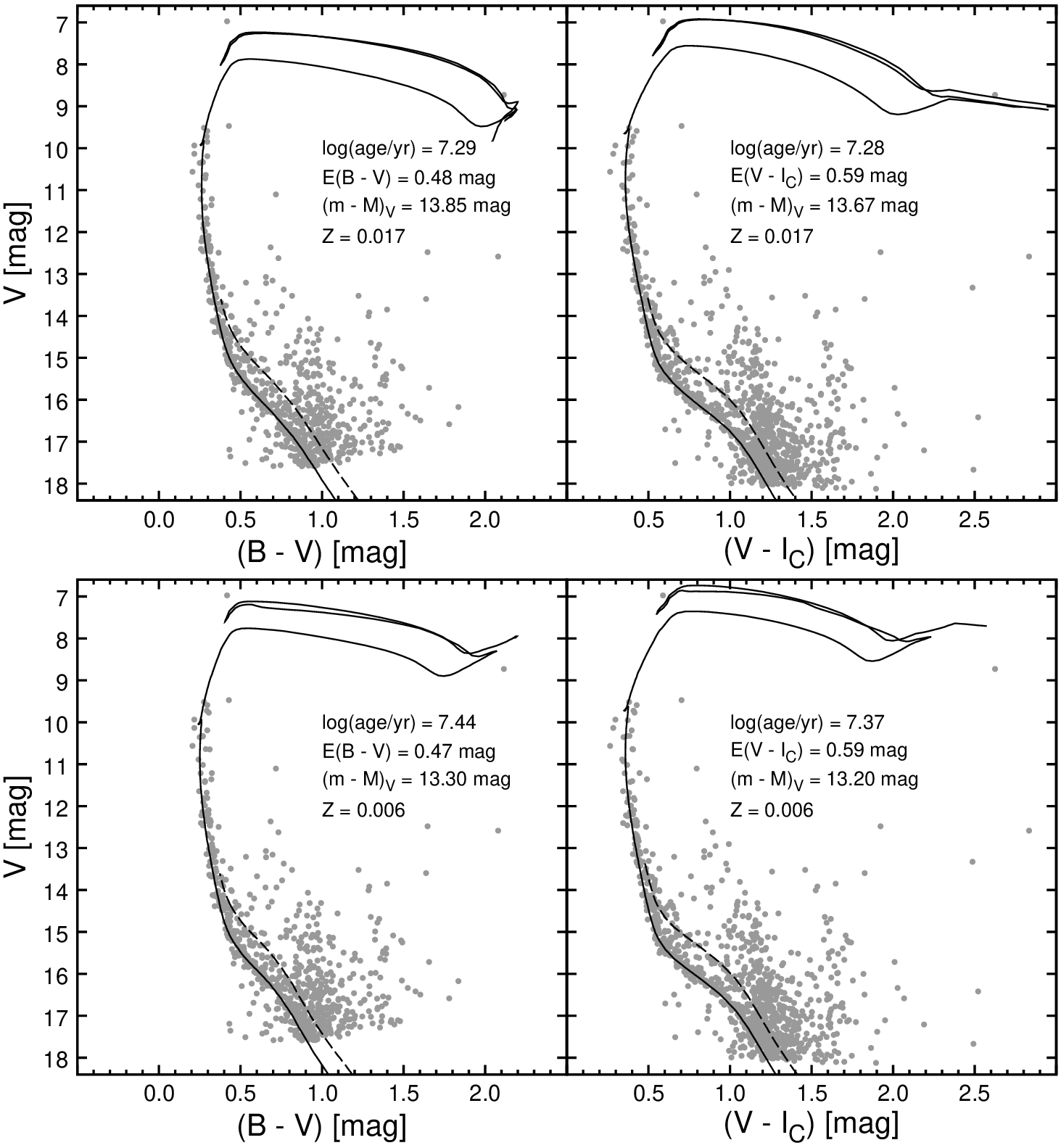}
\FigCap{$V$ vs. $(B-V)$ (left panels) and $V$ vs. $(V-I_{\textup{C}})$ (right panels) CM diagrams for NGC 457. The two upper panels show the best fits for solar metallicity models ($Z=$~0.017), the lower panels, for the metallicity of $Z=$~0.006. The fitted isochrones were taken from Bertelli {\etal}(2008, 2009). The resulting parameters are shown in all panels. In each panel the dashed line is a fragment of the isochrone shifted upward by 0.754~mag, showing the sequence of member equal-mass binaries.}
\label{cmd}
\end{figure}

The main sequence of the cluster is clearly visible in both CM diagrams. It is narrow, implying a small differential reddening across the cluster. In the lower part, a contribution from field stars becomes apparent. There is also a sequence of stars parallel to the lower part of the cluster main sequence, which is probably a binary sequence observed also in some other open clusters (e.g.~Bolte 1991). Its existence in NGC\,457 was already noted by Phelps and Janes (1994).

In order to derive the parameters of the cluster, we fitted isochrones from Bertelli {\etal}(2008, 2009) to the cluster main-sequences independently in the $V$ {\it vs.}~$B-V$ and in the $V$ {\it vs.}~$V-I_{\textup{C}}$ CM diagrams. The first fit (upper left panel of Fig.\,2) results in $\log(\mbox{age/yr}) =$ 7.29, $(m-M)_{\textup{V}}=$ 13.85~mag, and $E(B-V)=$ 0.48~mag. From the other one (upper right panel of Fig.\,2), we got $\log(\mbox{age/yr}) =$ 7.28, $(m-M)_{\textup{V}}=$ 13.67~mag, and $E(V-I_{\textup{C}})=$ 0.59~mag. In both cases we adopted models with solar metallicity, $Z=$ 0.017. Whether the metallicity of the cluster is indeed close to solar needs to be verified spectroscopically. The only metallicities available for NGC\,457 are those of Tadross (2003), who derived [Fe/H] $= -$0.46, and Fitzpatrick and Massa (2007), who give [M/H] $=-$0.43. The former value was obtained from the $UBV$ photometry, the latter, from modeling spectral energy distribution of five cluster members in the range from 
ultraviolet to infrared. If the lower metallicity is assumed in accordance with these determinations, an isochrone with larger age, $\log(\mbox{age/yr}) \approx$ 7.4, fits the cluster main sequence best and the resulting apparent distance modulus $(m-M)_{\textup{V}}\approx$ 13.3~mag (Fig.~2, lower panels). Summarizing, we estimate the age of the cluster for 20 $\pm$ 5~Myr; the uncertainty includes also the possibility of the lower-than-solar metallicity of the cluster. As far as the distance modulus is concerned, we adopt the true distance modulus $(m-M)_0=$ 12.3 $\pm$ 0.2~mag corresponding to the distance of 2.9 $\pm$ 0.3 kpc. If the metallicity of the cluster is assumed to be equal to $Z=$ 0.006, the distance to the cluster is smaller and amounts to about 2.3~kpc.

In addition to the transformation of the photometry, we also made astrometric transformation of average stellar positions to equatorial coordinates. We used the UCAC3 catalog (Zacharias {\etal}2010) as a reference. Standard deviation of this transformation amounted to about 0.1$^{\prime\prime}$. The table with positions of stars and our $BVI_{\rm C}$ and H$\alpha$ photometry is available from the \textit{Acta Astronomica Archive}.

\section{H$\alpha$ Photometry and Be Stars}
There were several searches for Be stars in open clusters that included NGC\,457 as a target. The emission was detected either by inspecting spectrograms (also objective prism spectrograms) that included H$\alpha$ (Schild and Romanishin 1976, Kohoutek and Wehmeyer 1997, Torrej\'on {\etal}1997, Mathew {\etal}2008) or by means of the narrow-band H$\alpha$ photometry. The results of these searches are summarized in Table 1. The papers which are the sources of the information on emission are abbreviated in the header; the abbreviations can be found in the references. Unfortunately, although the above studies included many cluster members, the authors rarely documented non-detections. For this reason, the ellipsis in a column stands both for the lack of observation and for the lack of information if a star was observed. Detections of emission are marked by `$+$', non-detections, by `$-$' (less certain if followed by a colon). There are two entries for the present paper separated by `/'; the first stands for the 
emission measured from the 1999\,--\,2002 observations, the other one, from those in 2013. In addition, some spectral types for stars in the field of NGC\,457 published in the literature and indicating the presence of Balmer line emission are given in the last column of Table 1. The table contains data for 15 stars; all but one are the cluster members. This makes NGC\,457 quite rich in Be stars.
\MakeTable{rrccccccl}{\textwidth}{Be stars in the field of NGC\,457\label{Be}}
{\hline\noalign{\smallskip}
&&\multicolumn{5}{c}{Source of information on emission} & &\\
Star & $V$ & SR76 & KW97 & T97 & M08 &This paper & $\Delta\alpha$ & Spectral type(s), remarks\\
\noalign{\smallskip}\hline\noalign{\smallskip}
131 &   6.97 & \ldots & \ldots         & \ldots & \ldots & {\ldots} / $-:$ & {\ldots}& B5e$\alpha$ [1], B1e$\alpha$ [2], \\
&&&&&&&&B6\,Ib [3,4,5], B5-6 Ib [6],\\
&&&&&&&&B5\,Ib [7]\\
153 &   9.47 & \ldots & $+$        & $+$    & \ldots & {\ldots} / $+$ &{\ldots}&B0:\,IV:e [3]\\
198 &   9.59 & \ldots & \ldots       & $+$    & \ldots & {\ldots} / $+$ &{\ldots}&B1.5\,V:pe [3]\\
128 &   9.72 &  $+$  & $+$        & $+$    & \ldots & $-$ / $-$    & $+$0.013 &B6\,V:(e) [8],\\
&&&&&&&&fast-rotating star\\
143 &   9.95 & \ldots & $+$       & \ldots & \ldots & $+$ / $+$    & $+$0.070 & foreground Be star,\\
&&&&&&&&fast-rotating star [11]\\
  14 & 10.35 & $+$   & $+$        & $+$    & \ldots & $+$ / $+$   & $-$0.249 & B2:e [9]\\
 100 & 10.64 & \ldots & \ldots     & \ldots & \ldots & {\ldots} / $+:$ &{\ldots}& \\
 154 & 11.21 & \ldots & \ldots     & \ldots & \ldots & {\ldots} / $+:$ &{\ldots}& \\
    6 & 11.36 & \ldots & \ldots     & \ldots  & $+$   & $+$ / $+$  & $-$0.297 & B3\,Ve [10], \\
&&&&&&&&Be star in shell phase\\
  91 & 11.40 & $+$   & \ldots     & $+$    & \ldots & $+$ / $+$   & $-$0.102 & \\
  62 & 11.81 & \ldots & \ldots     & \ldots & \ldots &  $-$ / $+$ & $-$0.112 &\\
 124 & 11.99 & \ldots & \ldots     & \ldots &  $+$   & {\ldots}  / $+$  & {\ldots}&B3\,Ve [10]\\
 117 & 12.66 & \ldots & \ldots     & \ldots & \ldots &  $-$ / $+$ & $-$0.240 &\\
  87 & 12.91 & \ldots & \ldots     & \ldots & \ldots &  $+$ / $-$ & $+$0.106 &\\
  43 & 12.98 & \ldots & \ldots     & \ldots & \ldots &  $+$ / $-$ &$+$0.166 & fast-rotating star,\\
  &&&&&&&&possible binary [11]\\
  \noalign{\smallskip}\hline\noalign{\smallskip}
\multicolumn{9}{l}{%
  \begin{minipage}{0.95\textwidth}%
    \scriptsize References to the entries in the last column: [1] -- Merrill (1935), [2] -- O'Keefe (1941), [3] -- Morgan {\etal}(1955), [4] -- Hiltner (1956), [5] -- Burnichon (1975), [6] -- Berger (1962), [7] -- Lennon {\etal}(1992), [8] -- Hoag and Applequist (1965), [9] -- Miller and Merrill (1951), [10] -- Mathew and Subramaniam (2011), [11] -- Huang and Gies (2006a).
  \end{minipage}%
}\\
}

\begin{figure}[ht]
\includegraphics[angle=0,width=0.92\textwidth]{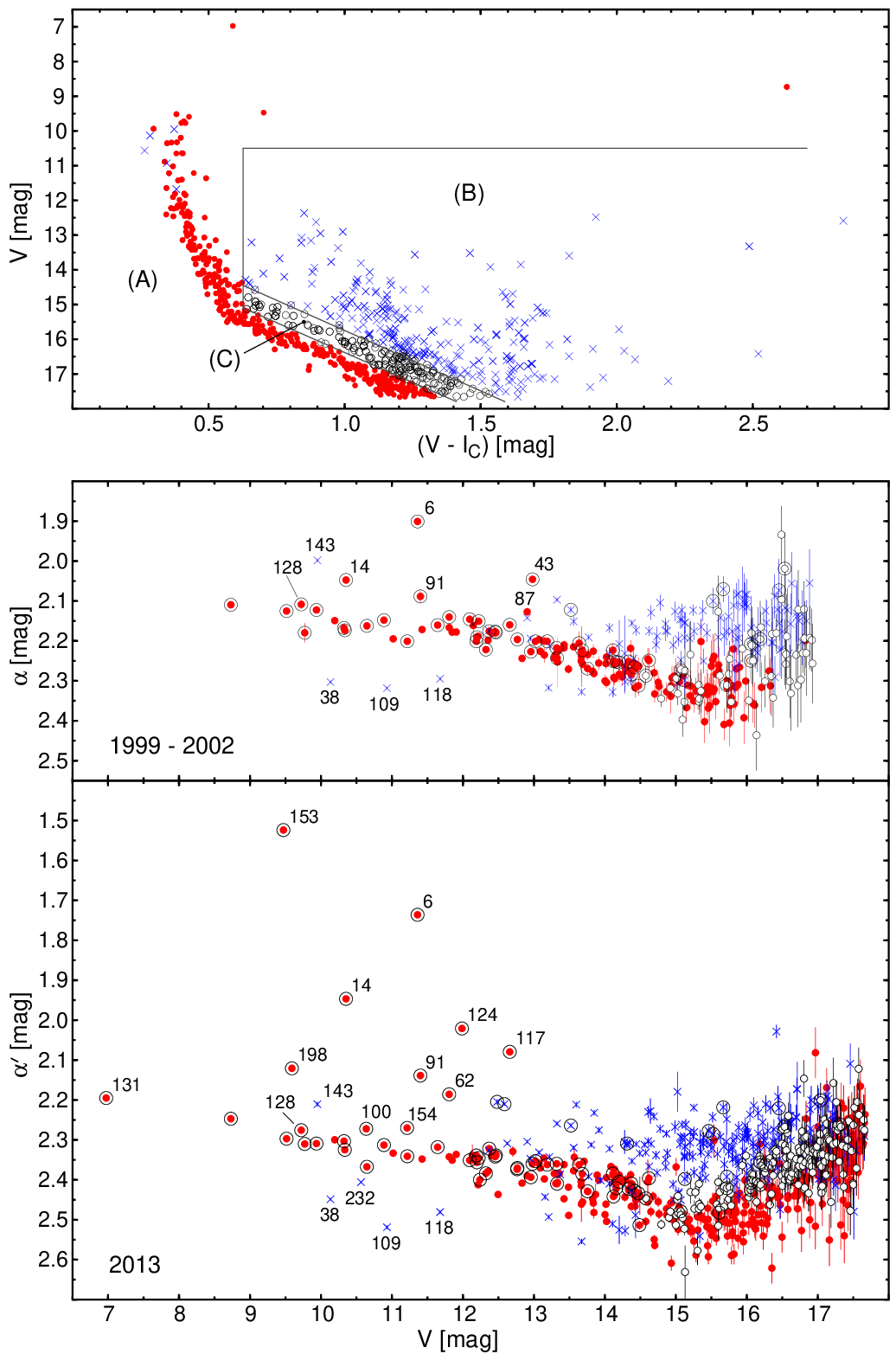}
\FigCap{\textit{Top:} Color-magnitude diagram for stars with measured $\alpha$ and $\alpha^\prime$ indices, see text for explanation. In all panels, stars belonging to groups A, B, and C are plotted with dots, crosses and open circles, respectively. Explanation of the division of stars into the three groups is given in the text. \textit{Bottom:} The $\alpha$ and $\alpha^\prime$ indices plotted as a function of the $V$ magnitude for stars in the field of NGC\,457. Symbols for variable stars are encircled. Be stars and four bright non-members are labeled with their WEBDA numbers.
}
\label{alpha}
\end{figure}

Our H$\alpha$ observations were carried out twice: during the second variability run on ten nights between December 4, 1999 and February 16, 2002 and as a part of the ongoing H$\alpha$ survey of open clusters, on July 25 and 27, 2013. In both cases, a pair of H$\alpha$ filters, narrow and wide, was used. The narrow filters in both instrumental systems were similar having 3-nm full-width-at-half-maximum (FWHM) transparency curves, while the wide filters were different. That used in the years 1999\,--\,2002 had 20-nm FWHM, whereas the new one, used in 2013, was wider, with FWHM equal to 40~nm. 

For each of the five fields in NGC\,457 observed during the second run (1999\,--\,2002), we obtained 6 to 8 narrow-band and 8 to 13 wide-band images, depending on the field. An average integration time amounted to about 37 minutes in narrow and 6 minutes in the wide filter. In 2013, the integration times were equal to 25 minutes in the narrow and 5 minutes in the wide filter. The 2013 H$\alpha$ observations were carried out in three overlapping fields and covered all stars observed during all three runs of our variability survey. In addition, the 2013 H$\alpha$ photometry was deeper than the 1999\,--\,2002 one. This resulted in a larger number of stars with measured H$\alpha$ index in 2013 (757 stars) than in 1999\,--\,2002 (316 stars). In both cases, we limited the number of stars to those with the uncertainties of the $\alpha$ or $\alpha^\prime$ index (see below for the definition) smaller than 0.07~mag.

The resulting reddening-free $\alpha$ index, defined as a difference between the magnitude in the narrow and wide filter (Pigulski {\etal}1997), is a measure of the equivalent width of the H$\alpha$ line and allows detecting H$\alpha$ emission in all B-type stars in NGC\,457.  Although the $\alpha$ index is defined in the same way for the old and new instrumental systems, the indices are not directly comparable because the filters are different. For this reason, the old and new index will be referred to as $\alpha$ and $\alpha^\prime$, respectively (see Fig.~3). Using data for 132 stars with $V<$~15 mag without emission in H$\alpha$ which have both $\alpha$ and $\alpha^\prime$ measured, we derived the following relation between the two indices:
\begin{equation}
\alpha^\prime = (1.107\pm 0.043)\times (\alpha - 2.2) + (2.3642\pm 0.0023).
\end{equation}
The relation is shown in Fig.~4a. The r.m.s.~of residuals from the fit amounts to 0.025~mag.
\begin{figure}[ht]
\includegraphics[angle=0,width=\textwidth]{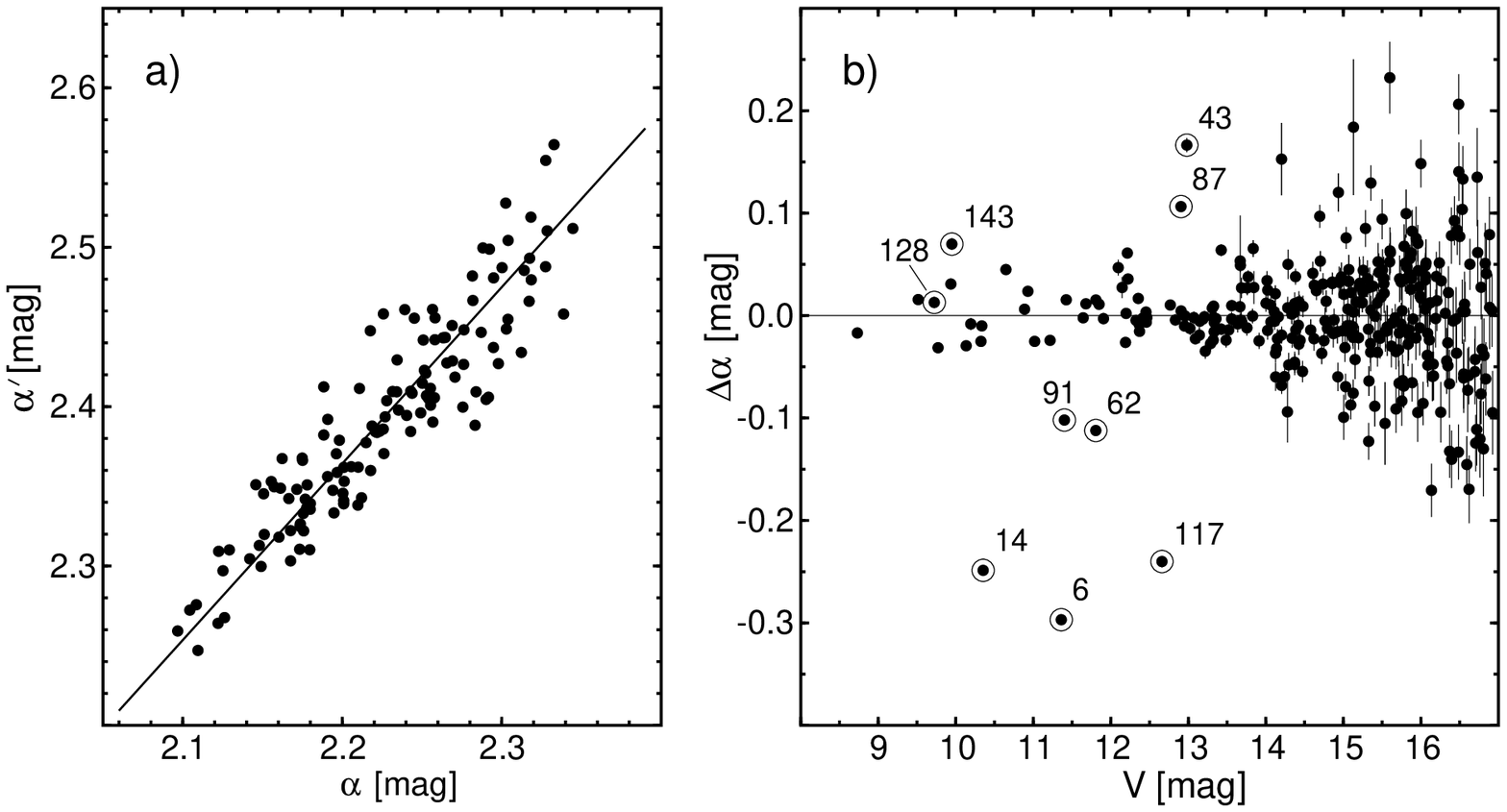}
\FigCap{a) Dependence between the two instrumental H$\alpha$ indices, $\alpha$ and $\alpha^\prime$, see text for explanation. Solid line is the regression line defined by Eq.~(4). b) The difference $\Delta\alpha$ plotted as a function of $V$ magnitudes for all stars having both $\alpha$ and $\alpha^\prime$ measured. Symbols for Be stars are encircled and labeled with the WEBDA numbers.}
\label{alpha}
\end{figure}

The location of stars in Fig.\,3 can be understood when the observed values of $\alpha$ are compared with the synthetic ones. This is shown by Ko{\l}aczkowski {\etal}(2004) in their Fig.\,8. The main-sequence members of the cluster without emission in H$\alpha$ form a well-defined sequence with $\alpha$ (or $\alpha^\prime$) increasing when going from the earliest spectral types towards spectral types A0-A1 (at $V$ magnitude $\sim$15.5 for NGC\,457), then decreasing for later types (lower panel of Fig.\,3). In the range covered by the B-type stars, we can easily indicate Be stars as well as some, mostly A-type, non-members. The former will have $\alpha$ (or $\alpha^\prime$) smaller than the cluster members without emission, the latter, higher. In this way, four bright stars, NGC\,457-38, -109, -118 and -232,
could be identified as non-members despite the fact that they fit the cluster main sequence in the CM diagram quite well (see the upper panel of Fig.\,3). It is worth noting that these four stars were already identified as non-members by Pesch (1959) based on his $UBV$ photometry. A comparison of the two bottom panels of Fig.\,3 shows also that for a given $V$ magnitude $\alpha^\prime$ is more accurate than $\alpha$. Consequently, the following discussion of H$\alpha$ emission in the members of NGC\,457 will be based mainly on the $\alpha^\prime$ vs.~$V$ plot. 

It can be seen from Fig.\,3 (bottom panels) that there are 11 B-type stars with clear emission in H$\alpha$, of which 10 are cluster members (NGC\,457-6, -14, -43, -62, -87, -91, -117, -124, -153 and -198). The remaining star is NGC\,457-143 (BD\,+57$^\circ$263). It is a B-type star with H$\alpha$ in emission. According to Huang and Gies (2006b), its effective temperature amounts to about 12\,600\,K, i.e.~corresponds to spectral type B7 (Gray and Corbally 1994). With $V=$ 9.95~mag, the star is too bright to be a cluster member of this type. Pesch (1959) classified the star as a non-member because it was much less reddened than cluster members. We therefore conclude that NGC\,457-143 is a foreground Be star. The strongest emission (i.e. the smallest $\alpha^\prime$) has been detected in NGC\,457-153 (BD\,+57$^\circ$243) and NGC\,457-6. Surprisingly, the latter star turned out to be in a shell phase (see Sect.~5). Given the uncertainties of $\alpha^\prime$ we may conclude that two other stars, NGC\,457-100 and 
-154, show probably a weak emission.

Six Be stars in the field are new detections. Two (NGC\,457-43 and -87) show emission in 1999\,--\,2002 observations, but their $\alpha^\prime$ do not indicate the emission in 2013. For NGC\,457-43, Huang and Gies (2006a) derived the projected rotational velocity $v\sin i=$ 389 $\pm$ 11 km\,s$^{-1}$ and based on two discrepant values of radial velocity classified the star as a single-lined spectroscopic binary. They noted, however, that He\,{\sc i} lines suggest large $v\sin i$, while H$\gamma$ does not. It is possible that the large $v\sin i$ is the result of unresolved line-splitting due to orbital motion. The star shows clear emission, however, and therefore is a likely Be star in a double system. In two other stars, NGC\,457-62 and -117, the opposite behavior is observed: they show no emission in the 1999\,--\,2002 data and clear emission in 2013. We conclude that the four newly detected Be stars in NGC\,457 are examples of switching between Be and non-Be phases observed in Be stars. There is another 
example of this kind of switching among the members of NGC\,457. Star NGC\,457-128 (MWC\,423, BD\,+57$^\circ$246) was clearly a Be star in the past (see Table 1), but our H$\alpha$ photometry does not indicate emission. It is thus possible that at the two epochs of our observations, the emission dropped to the level that could not be detected by means of our H$\alpha$ photometry. The spectroscopic observations described in the next section show, however, that NGC\,457-128 is a fast-rotating star. This fact is consistent with the above picture of a non-Be phase of a Be star. Finally, we need to comment on the H$\alpha$ emission in NGC\,457-131 (HD\,7902, B6\,Ib), the brightest star we observed. The emission in this star was suggested in the past (Table 1). However, this is a B-type supergiant and therefore should have smaller $\alpha$. The $\alpha^\prime$ value of 2.1951 $\pm$ 0.0006 we measured for this star in 2013 shows that although a weak emission in H$\alpha$ cannot be excluded, there is no clear 
evidence for its presence.

The region we observed is contaminated by field stars, mostly with $V >$ 12 mag (Fig.\,2). In order to see how these stars contaminate the $\alpha$ (and $\alpha^\prime$) vs.~$V$ diagram, we plotted a CM diagram for stars having measured $\alpha$ and $\alpha^\prime$ indices (upper panel of Fig.\,3) and divided the stars into the following three groups: (A) likely cluster members, (B) non-members redder than the main sequence stars in the cluster, (C) the possible double-star sequence of cluster stars. The three groups were plotted with different symbols in the $\alpha$ -- $V$ and $\alpha^\prime$ -- $V$ diagrams. As expected, the non-members (group B) have mostly smaller $\alpha$ (or $\alpha^\prime$) than cluster stars of the same magnitude. It is interesting to note that stars from group C (suspected double-star sequence) fall mostly in the same place where the cluster members do. This is a strong argument in favor of the cluster membership of many of these stars and, consequently, the presence of the double-
star sequence in the cluster.

The fact that the two epochs of the H$\alpha$ observations are separated by more than 10 years allows us to discuss changes of emission in B-type stars. Such changes are quite common among Be stars. It is also known that Be stars can switch between three different phases: a normal B-type (non-Be) star, Be star, and a shell Be star (see, e.g.~Porter and Rivinius 2003 and references therein). Assuming that the relation between $\alpha$ and $\alpha^\prime$ (Fig.~4a) is valid also beyond the range used to derive Eq.~(4), we defined $\Delta\alpha = \alpha^\prime\mbox{(observed)} - \alpha^\prime\mbox{(transformed)}$, where $\alpha^\prime\mbox{(transformed)}$ represents the $\alpha$ measured in the 1999\,--\,2002 and transformed to $\alpha^\prime$ according to Eq.~(4). The result is shown in Fig.~4b. It can be seen from this figure that for five Be stars (NGC\,457-6, -14, -62, -91, and -117) the H$\alpha$ emission is stronger in 2013 than in 1999\,--\,2002 ($\Delta\alpha<0$), for the other three (NGC\,457-43, -87, -
143), the opposite is true ($\Delta\alpha>0$), while in NGC\,457-128 the emission, if present at all (see above), remained virtually unchanged.

\begin{figure}[!th]
\includegraphics[angle=0,width=12cm]{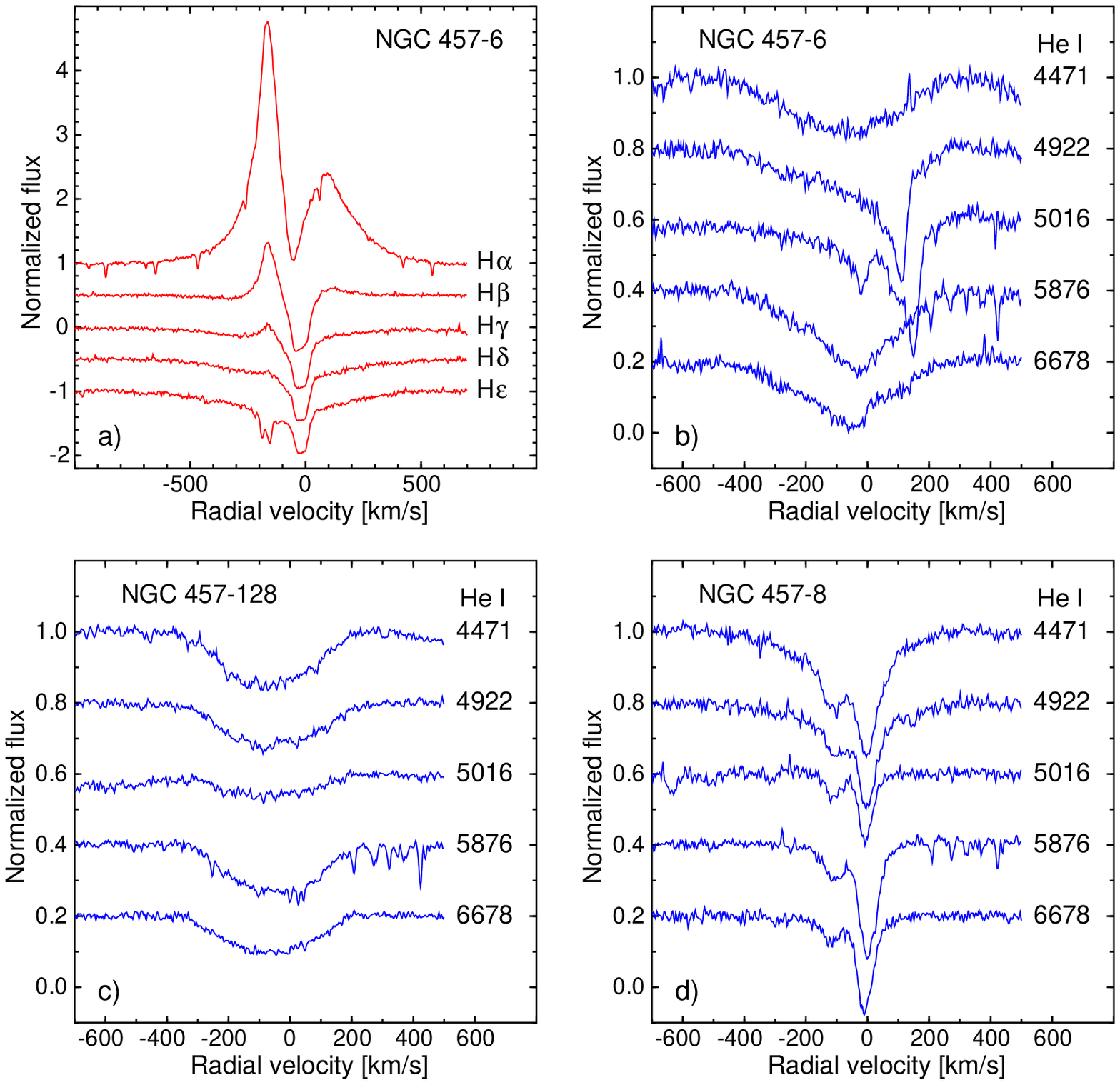}
\FigCap{Selected Balmer series lines and He\,{\sc i} lines in the APO spectra of three stars in NGC\,457: NGC\,457-6, the Be star in shell phase (two upper panels), NGC\,457-128, Be star in a non-emission phase (lower left) and NGC\,457-8, $\beta$~Cep-type star (lower right). Note line doubling in NGC\,457-8. Strong absorption lines superimposed on He\,{\sc i} 4922 and 5016\,{\AA} lines of NGC\,457-6 are shell Fe\,{\sc ii} lines. For clarity, the lines are shifted in flux by 0.5 (Balmer series) or 0.2 (He lines).}
\label{alpha}
\end{figure}

\section{The APO Spectroscopy}
As a part of one of our spectroscopic programs, single-epoch spectra of three stars in NGC\,457, NGC\,457-6 (Be star), -8 ($\beta$~Cep-type star) and -128 (Be star in a quiescent phase), were taken by one of us (DM) with the Apache Point Observatory (APO) ARC 3.5-m telescope and the ARC Echelle Spectrograph (ARCES) on two nights, August 13 and September 2, 2013. The spectra have a resolving power of 31\,500 and cover a range between 3200 and 10\,000~{\AA}. In order to derive the effective temperatures, $T_{\rm eff}$, and projected rotational velocities, $v\sin i$, we fitted the spectra using the BSTAR2006 grid of non-LTE model atmospheres of Lanz and Hubeny (2007) and the ROTIN3 program.

The following conclusions can be drawn from the APO spectra: (i) NGC\,457-6 is a Be star in a shell phase. The strongest emission can be seen in H$\alpha$, H$\beta$ (Fig.\,5a) and the oxygen O\,{\sc i} 8446\,{\AA} line. The shell absorption lines occur throughout the observed spectrum, but those of hydrogen are especially prominent. The lines of the Balmer series can be seen up to H32, those of the Paschen series, up to Pa30. The spectrum is very similar to phase B spectrum of $\zeta$~Tau shown in Fig.\,3 of \v{S}tefl {\etal}(2009). The $v\sin i$ amounts to 250 $\pm$ 30 km\,s$^{-1}$, $T_{\rm eff} =$ 16\,000~K. (ii) NGC\,457-128 is presently a normal fast-rotating B-type star (Fig.\,5c) showing no emission in Balmer lines. Given its previous Be character found by several authors (Table 1), it is a Be star presently in a non-emission phase. Its $v\sin i$ amounts to 232 $\pm$ 5 km\,s$^{-1}$, $T_{\rm eff} =$ 24\,000~K. (iii) NGC\,457-8, the only $\beta$~Cep-type variable in the cluster, is a binary. This can be 
inferred from line doubling seen very well in the He\,{\sc i} lines (Fig.\,5d). The mean separation of the lines of the components derived from the He\,{\sc i} lines amounts to 110~km\,s$^{-1}$. If we assume that the spectrum was obtained close to quadrature, the orbit is circular and the total mass of the binary amounts to 15\,$M_\odot$, we get only a very rough estimate of the maximum orbital period, viz., about 90 days. The true orbital period is probably much shorter. Both components of the binary rotate relatively slowly: we derived $v\sin i =$\,34\,$\pm$\,2 and 52\,$\pm$\,4 km\,s$^{-1}$ for the brighter (i.e., with stronger lines) and fainter component, respectively. The $T_{\rm eff}$ of the components was estimated to be 23\,000 and 20\,000\,K, respectively.

\section{Variable Stars}
NGC\,457 has been already a target of two variability searches. Using Schmidt telescopes Maciejewski {\etal}(2008, hereafter Mac08) observed a large field around the cluster covering the whole 30$^\prime\times$30$^\prime$ area shown in Fig.~1. They found 31 variables and concluded that six of them are true or likely members of NGC\,457. Recently, Zhang {\etal}(2012, hereafter Zha12) carried out another variability search in the cluster finding 13 variable stars. In addition, several stars in the field were claimed in the literature to be variable. In total, 28 stars in the field shown in Fig.\,1 were already known or were suspected to be variable. Out of the 28, 13 are located in the field covered during our observing runs. These variable stars are listed in Table 2. We confirm variability of all these stars.
\begin{figure}[ht]
\begin{center}
\includegraphics[width=0.9\textwidth]{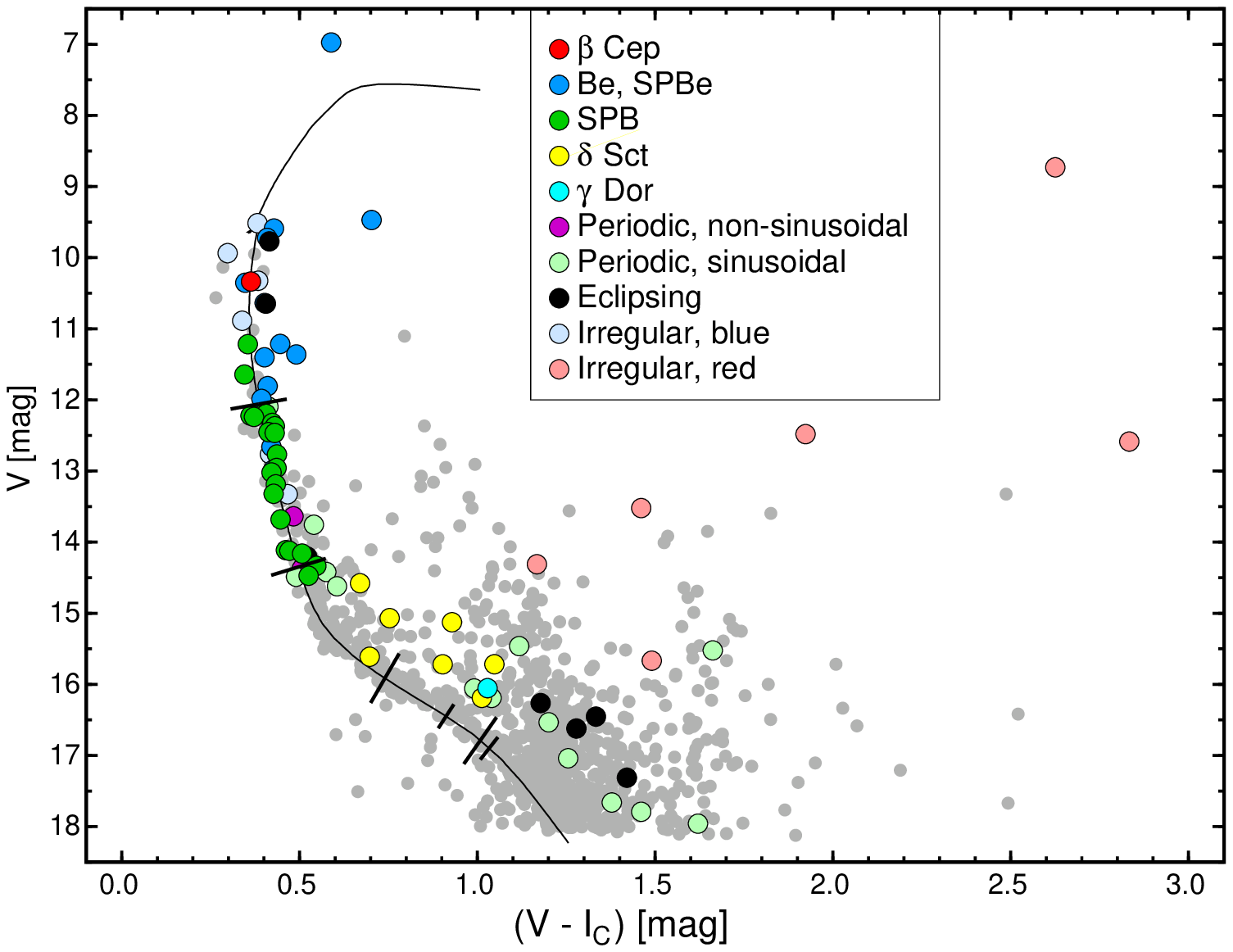}
\end{center}
\FigCap{Variable stars in the CM diagram of NGC\,457. The solid line is the same isochrone as in Fig.\,2. The short lines perpendicular to the isochrone mark (from top to bottom): faint limit of the $\beta$~Cep instability strip, faint limit of the SPB instability strip (both adopted from Miglio {\etal}2007), blue and red borders of the $\delta$~Sct and $\gamma$~Dor instability strips adopted from Dupret {\etal}(2004).}
\end{figure}
\begin{figure}[ht]
\begin{center}
\includegraphics[width=0.9\textwidth]{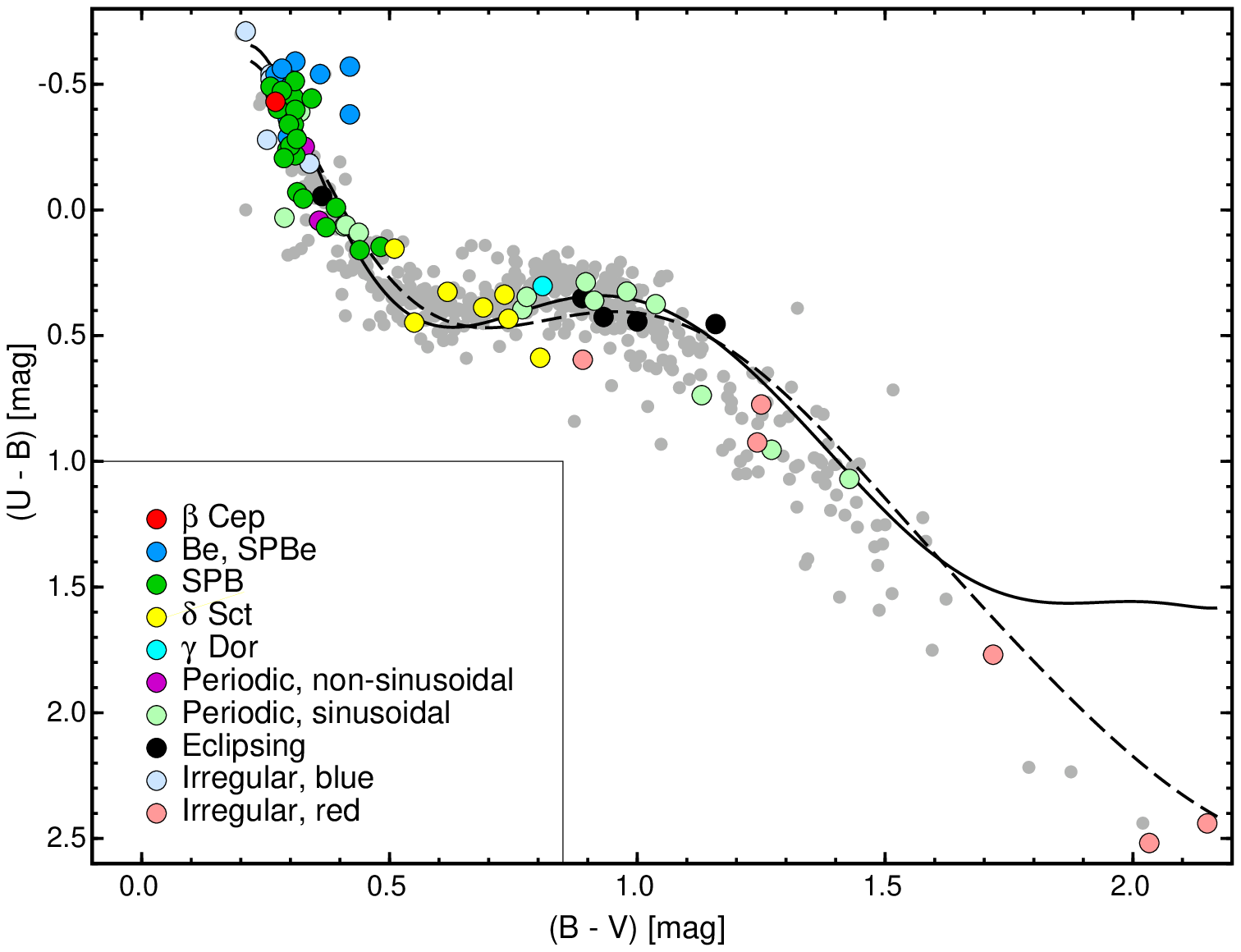}
\end{center}
\FigCap{The $(U-B)$ vs.~$(B-V)$ diagram for NGC\,457 with the positions of variables marked with the same symbols as in Fig.~6. The indices were taken from papers of Phelps and Janes (1994), Pesch (1959), Hoag {\etal}(1961) and Moffat and Vogt (1974). The two lines correspond to the intrinsic luminosity class V (continuous line) and I (dashed line) relations from Caldwell {\etal}(1993) shifted adopting $E(B-V)=$~0.48~mag and $E(U-B)/E(B-V)=$~0.72.} 
\end{figure}
\begin{figure}[ht]
\begin{center}
\includegraphics[angle=270,width=0.8\textwidth]{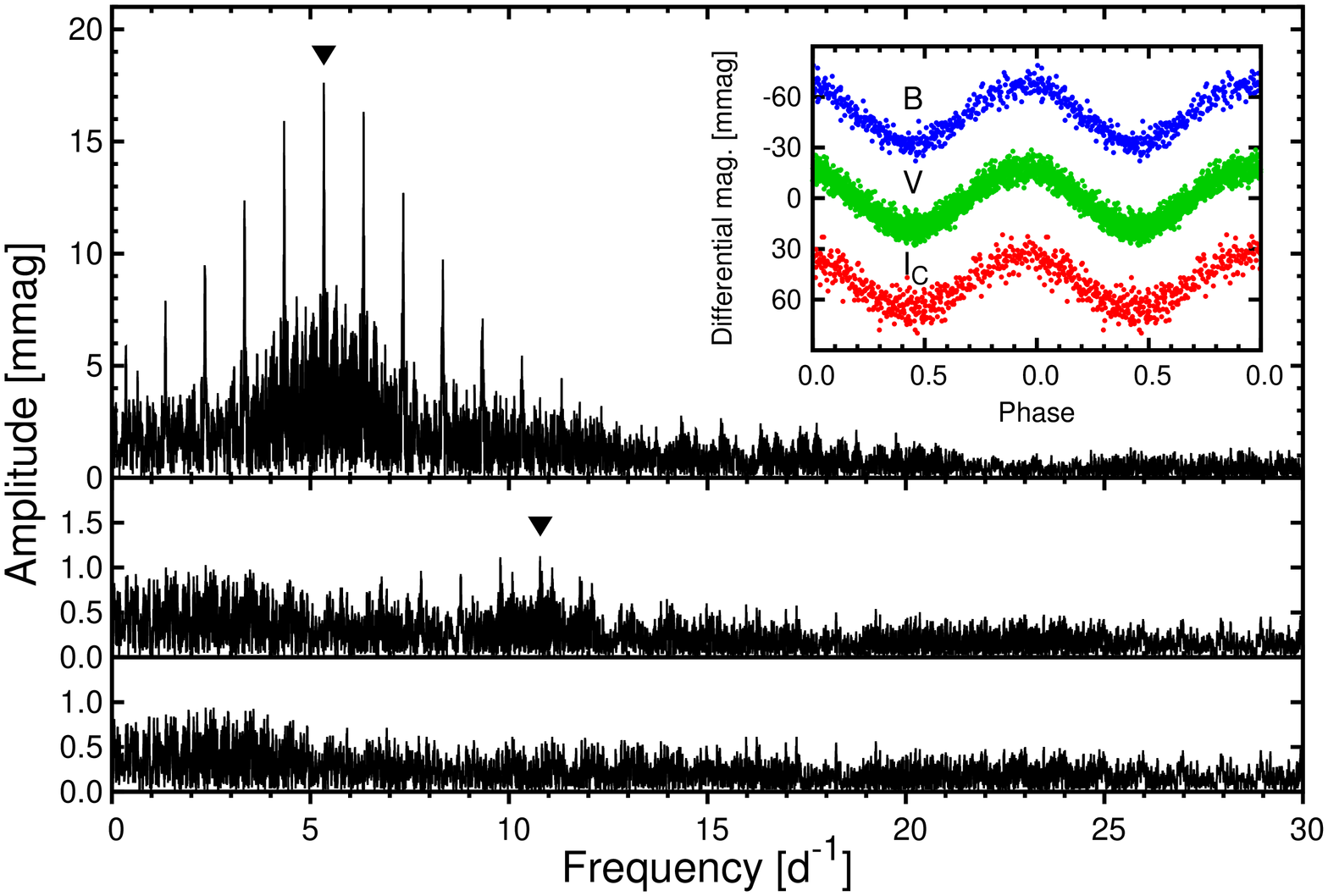}
\end{center}
\FigCap{Fourier amplitude spectra of the combined $V$-filter photometry of NGC457-8 and -6661: original data (top), after removing $f_1=$ 5.33987~d$^{-1}$ (middle), and after removing $f_1$ and $f_2=$ 10.7856~d$^{-1}$ (bottom). The inset shows $BVI_{\rm C}$ data freed from $f_2$ and folded with $P_1=f_1^{-1}$.}
\end{figure}

\MakeTable{rccrcccl}{\textwidth}{Previously known variable stars in the field covered in this study}
{\hline\noalign{\smallskip}
Star & $\alpha_{\rm 2000.0}$ & $\delta_{\rm 2000.0}$ & $V$ & Mac08 & Zha12 & Var.~type & Other name\\
\noalign{\smallskip}\hline\noalign{\smallskip}
  131 & 1:19:51.74 & +58:12:29.3 &   6.97 & --- & --- & Be & NSV~466 \\
    25 & 1:19:53.52 & +58:18:33.8 &   8.73 & V1 & --- & Irr (Red) & V466~Cas \\
  153 & 1:18:54.11 & +58:12:07.2 &   9.47 & --- & --- & Be & NSV\,463 \\
  128 & 1:19:08.56 & +58:14:16.2 &   9.72 & --- & v1 & Be &\\
    37 & 1:19:22.96 & +58:18:21.6 &   9.77 & --- & v2   & EA & \\
      8 & 1:19:28.22 & +58:17:19.4 & 10.34 & --- & v4   & $\beta$~Cep & \\
    14 & 1:19:31.21 & +58:15:51.5 & 10.35 & --- & v3 & Be & \\
    85 & 1:19:09.01 & +58:17:26.8 & 10.65 & V3 & v5  & EA & V765 Cas \\
    91 & 1:19:14.46 & +58:13:34.1 & 11.40 & --- & v6 & Be &  \\
  124 & 1:19:02.32 & +58:19:20.8 & 11.99 & V5 & v7 & Be & V1086 Cas \\
    52 & 1:19:36.78 & +58:15:04.6 & 12.37 & --- & v8 & SPB & \\
3318 & 1:19:23.08 & +58:16:02.1 & 14.58 & --- & v11 & $\delta$~Sct/$\gamma$~Dor & \\
3328 & 1:19:29.46 & +58:13:42.2 & 16.54 & V2 & --- &  EW & V1090 Cas \\
\noalign{\smallskip}\hline
}

The search for variables was done independently for all three runs. For obvious reasons, the observations made during the third run represented the best data set both in the sense of quality and sky coverage. They were the subject of the most thorough variability search. As a result, we have found 79 variable stars. Since 13 were already known (Table 2), 66 are new discoveries. A preliminary report from this work, including the discovery of the $\beta$~Cep-type pulsations in NGC\,457-8, has been already published by Mo\'zdzierski {\etal}(2013).

The variability search was based mainly on the Fourier periodograms which were calculated up to 80 d$^{-1}$. In addition, the light curves, periodograms and phased light curves for the brightest stars were inspected by eye. For the remaining stars, this was done only for the objects that showed an excess signal in Fourier amplitude spectra. The stars were also searched for multiperiodicity using the standard prewhitening procedure. The variable stars we found are listed in Table 3. Their location in the CM and color-color diagrams is shown in Figs.\,6 and 7. Whenever possible, they were assigned a variability class based on the light curve characteristics, period(s) and other properties such as spectral types or $UBV$ photometry.

\MakeOwnTable{rccrrlcl}{\textwidth}{Variable stars in NGC\,457}{\tiny}
{\hline\noalign{\smallskip}
      & $\alpha_{\rm 2000.0}$ & $\delta_{\rm 2000.0}$ & \multicolumn{1}{c}{$V$} &\multicolumn{1}{c}{$V-I_{\rm C}$} &\multicolumn{1}{c}{$\Delta V$}& Var.& \\
 Star & [$^{\rm h}$~~~$^{\rm m}$~~~$^{\rm s}$]&[$^{\rm o}$~~~$^\prime$~~~$^{\prime\prime}$] & \multicolumn{1}{c}{[mag]}& \multicolumn{1}{c}{[mag]} & \multicolumn{1}{c}{[mag]} &  type & Remarks\\
\noalign{\smallskip}\hline\noalign{\smallskip}
  8 & 1 19 28.27 & +58 17 18.1 & 10.34 & 0.36 & 0.034 & $\beta$~Cep & v4 (Zha12)\\
\noalign{\smallskip}\hline\noalign{\smallskip}
 131 & 1 19 51.72 & +58 12 29.5 &  6.97 & 0.59 & 0.09 & Be: & HD\,7902, NSV 466\\
 153 & 1 18 54.09 & +58 12 07.0 &  9.47 & 0.70 & 0.05 & Be & NSV\,463, MWC\,423 \\
 198 & 1 18 33.06 & +58 22 30.7 &  9.59 & 0.43 & 0.08 & Be & HD\,236689 \\
 128 & 1 19 08.56 & +58 14 15.7 &  9.72 & 0.41 & 0.05 & Be & BD$+$57$^{\rm o}$246, v1 (Zha12)\\
  14 & 1 19 31.25 & +58 15 50.0 & 10.35 & 0.35 & 0.04 & Be & BD$+$57$^{\rm o}$251, v3 (Zha12)\\
 100 & 1 19 46.41 & +58 12 26.1 & 10.64 & 0.40 & 0.08 & Be: & weak emission\\
 154 & 1 19 02.15 & +58 11 46.7 & 11.21 & 0.45 & 0.02 & Be: & weak emission\\
   6 & 1 19 32.99 & +58 17 25.7 & 11.36 & 0.49 & 0.13 & Be & shell star \\
  91 & 1 19 14.45 & +58 13 33.3 & 11.40 & 0.40 & 0.07 & Be & v6 (Zha12) \\
  62 & 1 19 56.86 & +58 15 57.4 & 11.81 & 0.41 & 0.005 & SPBe & \\
 124 & 1 19 02.35 & +58 19 20.4 & 11.99 & 0.39 & 0.05 & Be & V1086\,Cas, V5 (Mac08), v7 (Zha12) \\
 117 & 1 19 35.26 & +58 21 48.1 & 12.66 & 0.42 & 0.03 & Be & \\
  43 & 1 19 17.79 & +58 16 43.4 & 12.98 & 0.43 & 0.005 & SPBe & \\
\noalign{\smallskip}\hline\noalign{\smallskip}
  34 & 1 19 30.37 & +58 18 03.3 & 11.22 & 0.35 & 0.004 & SPB & \\
  71 & 1 19 44.63 & +58 20 30.6 & 11.64 & 0.35 & 0.004 & SPB & \\
   35 & 1 19 26.38 & +58 18 10.7 & 12.19 & 0.40 & 0.005 & SPB & \\
   60 & 1 19 50.48 & +58 15 56.6 & 12.20 & 0.41 & 0.006 & SPB & $\ast$ \\
  119 & 1 19 11.18 & +58 20 30.7 & 12.22 & 0.36 & 0.005 & SPB & \\
   84 & 1 19 04.18 & +58 17 49.0 & 12.24 & 0.37 &0.003 & SPB & \\
   64 & 1 19 56.18 & +58 16 23.3 & 12.32 & 0.42 & 0.006 & SPB & \\
   52 & 1 19 36.80 & +58 15 02.8 & 12.37 & 0.43 & 0.013 & SPB & $\ast$ v8 (Zha12) \\
   12 & 1 19 26.84 & +58 16 17.9 & 12.45 & 0.41 & 0.006 & SPB & \\
 116 & 1 19 46.52 & +58 21 19.8 & 12.46 & 0.43 &0.007 & SPB & $\ast$ \\
   48 & 1 19 26.11 & +58 15 48.5 & 12.77 & 0.44 &0.005 & SPB & $\ast$ \\
  32 & 1 19 35.43 & +58 19 01.9 & 12.96 & 0.44 & 0.008 & SPB & \\
  17 & 1 19 36.80 & +58 16 15.5 & 13.02 & 0.42 & 0.012 & SPB & $\ast$ \\
  42 & 1 19 21.35 & +58 16 48.3 & 13.19 & 0.43 & 0.010 & SPB & \\
  22 & 1 19 40.15 & +58 16 44.5 & 13.32 & 0.43 & 0.006 & SPB & $\ast$ \\
   3 & 1 19 40.82 & +58 17 36.7 & 13.68 & 0.45 &0.007 & SPB & $\ast$ \\
  20 & 1 19 40.73 & +58 16 09.7 & 14.11 & 0.46 & 0.010 & SPB & \\
  26 & 1 19 49.33 & +58 18 25.3 & 14.12 & 0.47 & 0.008 & SPB & $\ast$ \\
2270 & 1 19 48.16 & +58 18 35.8 & 14.16 & 0.51 & 0.016 & SPB & \\
   4 & 1 19 36.49 & +58 16 55.4 & 14.33 & 0.55 & 0.008 & SPB & $\ast$ \\
6166 & 1 19 41.30 & +58 17 27.4 & 14.47 & 0.53 & 0.004 & SPB & $\ast$ \\
\noalign{\smallskip}\hline\noalign{\smallskip}
3318 & 1 19 23.11 & +58 16 01.1 & 14.58 & 0.67 & 0.020 & $\delta$~Sct/$\gamma$~Dor & v11 (Zha12), non-member \\
1332 & 1 19 22.64 & +58 22 46.1 & 15.07 & 0.75 & 0.007 & $\delta$~Sct & \\
3233 & 1 19 37.35 & +58 15 41.9 & 15.13 & 0.93 & 0.009 & $\delta$~Sct & non-member\\
2380 & 1 19 00.96 & +58 20 48.8 & 15.61 & 0.70 & 0.007 & $\delta$~Sct & \\
3416 & 1 18 42.73 & +58 15 22.8 & 15.72 & 1.05 & 0.035 & $\delta$~Sct & non-member\\
3242 & 1 19 37.57 & +58 13 59.8 & 15.78 & 0.90 & 0.022 & $\delta$~Sct & \\
7202 & 1 19 12.41 & +58 15 12.6 & 16.19 & 1.01 & 0.018 & $\delta$~Sct & \\
\noalign{\smallskip}\hline\noalign{\smallskip}
2301 & 1 19 31.69 & +58 18 54.3 & 16.05 & 1.03 & 0.025 & $\gamma$~Dor & \\
\noalign{\smallskip}\hline\noalign{\smallskip}
 122 & 1 19 06.02 & +58 19 39.8 & 12.10 & 0.41 & 0.010 & Per & SPB candidate \\
  15 & 1 19 34.24 & +58 15 53.4 & 13.64 & 0.48 & 0.024 & Per &  CP star?, non-sinusoidal light curve \\
  93 & 1 19 28.48 & +58 13 56.4 & 13.76 & 0.54 & 0.005 & Per + Irr &  SPB candidate \\
  63 & 1 19 51.42 & +58 16 26.3 & 14.37 & 0.51 & 0.012 & Per &  Ell/Rot?, non-sinusoidal light curve \\
    2 & 1 19 42.93 & +58 17 27.7 & 14.42 & 0.58 & 0.004 & Per & SPB candidate \\
  126 & 1 18 52.99 & +58 16 46.9 & 14.49 & 0.49 & 0.010 & Per & SPB candidate\\
   58 & 1 19 45.38 & +58 14 39.3 & 14.62 & 0.61 & 0.010 & Per & SPB candidate, weak emission in H$\alpha$? \\
3421 & 1 18 45.45 & +58 14 00.1 & 15.46 & 1.12 & 0.012 & Per & non-member \\
3202 & 1 19 54.73 & +58 17 10.8 & 15.52 & 1.66 & 0.010 & Per & non-member \\
2218 & 1 19 41.45 & +58 22 08.6 & 16.06 & 0.99 & 0.012 & Per & \\
2424 & 1 18 40.44 & +58 18 53.1 & 16.19 & 1.04 & 0.017 & Per & \\
3349 & 1 19 08.83 & +58 13 45.4 & 16.53 & 1.20 & 0.035 & Per & non-member\\
6373 & 1 19 36.26 & +58 17 40.0 & 17.04 & 1.26 & 0.035 & Per & non-member\\
6697 & 1 19 27.97 & +58 15 08.7 & 17.66 & 1.38 & 0.053 & Per & non-member\\
6579 & 1 19 31.79 & +58 13 03.2 & 17.79 & 1.46 & 0.07 & Per & non-member\\
6464 & 1 19 34.41 & +58 14 52.9 & 17.96 & 1.62 & 0.15 & Per & non-member\\
\noalign{\smallskip}\hline\noalign{\smallskip}
  37 & 1 19 23.01 & +58 18 20.5 &  9.77 & 0.31 & 0.22 & EA & BD$+$57$^{\rm o}$249 (B1\,V), v2 (Zha12)\\
  85 & 1 19 09.04 & +58 17 26.3 & 10.65 & 0.31 & 0.55 & EA & V765\,Cas, V3 (Mac08), v5 (Zha12) \\
\noalign{\smallskip}\hline
}
  
\setcounter{table}{2}
\MakeOwnTable{rccrrlcl}{\textwidth}{Variable stars in NGC\,457 --- continued}{\tiny}
{\hline\noalign{\smallskip}
      & $\alpha_{\rm 2000.0}$ & $\delta_{\rm 2000.0}$ & \multicolumn{1}{c}{$V$} &\multicolumn{1}{c}{$V-I_{\rm C}$} &\multicolumn{1}{c}{$\Delta V$}& Var.& \\
 Star & [$^{\rm h}$~~~$^{\rm m}$~~~$^{\rm s}$]&[$^{\rm o}$~~~$^\prime$~~~$^{\prime\prime}$] & \multicolumn{1}{c}{[mag]}& \multicolumn{1}{c}{[mag]} & \multicolumn{1}{c}{[mag]} &  type & Remarks\\
 \noalign{\smallskip}\hline\noalign{\smallskip}
  11 & 1 19 32.95 & +58 16 35.0 & 14.20 & 0.52 & 0.013 & EA: & \\
2330 & 1 19 05.33 & +58 17 56.1 & 16.26 & 1.18 & 0.17 & EA & single eclipse, non-member\\
3328 & 1 19 29.46 & +58 13 40.8 & 16.54 & 1.33 & 0.24 & EW & V1090~Cas, V2 (Mac08), non-member \\
7723 & 1 18 55.72 & +58 17 45.3 & 16.63 & 1.26 & 0.36 & EA & non-member\\
7227 & 1 19 11.73 & +58 17 26.6 & 17.31 & 1.42 & 0.6 & EA & single eclipse, non-member \\
8306 & 1 18 31.57 & +58 19 28.0 & 18.84 & 1.31 & 0.3 & EW & \\
\noalign{\smallskip}\hline\noalign{\smallskip}
  19 & 1 19 34.17 & +58 15 22.4 &  9.52 & 0.38 & 0.03 & Irr (Blue) & BD$+$57$^{\rm o}$252 (B1\,IV) \\
 120 & 1 19 10.26 & +58 20 57.2 &  9.94 & 0.30 & 0.02 & Irr (Blue) & BD$+$57$^{\rm o}$247 (O9.5\,IV) \\
  33 & 1 19 35.78 & +58 19 15.5 & 10.32 & 0.38 & 0.03 & Irr (Blue) & BD$+$57$^{\rm o}$253 \\
  13 & 1 19 33.71 & +58 16 02.9 & 10.89 & 0.34 & 0.02 & Irr (Blue) & \\
 151 & 1 18 52.29 & +58 14 44.7 & 12.76 & 0.42 & 0.01 & Irr (Blue) & \\
  80 & 1 19 17.93 & +58 18 50.3 & 13.33 & 0.47 & 0.022 & Irr (Blue) & \\
\noalign{\smallskip}\hline\noalign{\smallskip}
  25 & 1 19 53.61 & +58 18 30.8 &  8.73 & 2.63 & 0.08 & Irr (Red) & HD\,236697 (M0.5 Iab), V466\,Cas, V1 (Mac08) \\
 130 & 1 19 37.25 & +58 11 38.5 & 12.48 & 1.92 & 0.02 & Irr (Red) & non-member\\
 199 & 1 18 28.21 & +58 18 37.9 & 12.59 & 2.83 & 0.10 & Irr (Red) & non-member\\
6798 & 1 19 25.11 & +58 15 58.1 & 13.52 & 1.46 & 0.02 & Irr (Red) & non-member\\
  150 & 1 18 41.34 & +58 16 52.4 & 14.31 & 1.17 & 0.04 & Irr (Red) & non-member\\
2338 & 1 19 12.64 & +58 19 14.7 & 15.67 & 1.49 & 0.045 & Irr (Red) & non-member\\
\noalign{\smallskip}\hline 
}

\subsection{$\beta$~Cep-Type Star NGC\,457-8}
The most luminous pulsating star in the cluster is NGC\,457-8, a $\beta$~Cep-type star discovered independently by Zha12 (their variable v4, see also Table 2). The star has no available spectral type, but its $UBV$ photometry (Moffat and Vogt 1974) and the APO spectrum (Sect.~5) indicate an early B-type. NGC\,457-8 has a fainter companion, NGC\,457-6661, located 4$^{\prime\prime}$ east of it. The stars are resolved in our images, but the companion affects the photometry of NGC\,457-8 leading to a significant noise at low frequencies. However, the noise is largely reduced if the fluxes from both stars are combined. The Fourier amplitude spectrum of the combined $V$-filter observations of NGC\,457-8 and -6661 is shown in Fig.~8. It reveals two significant peaks with frequencies of $f_1=$ 5.33987 and $f_2=$ 10.7856~d$^{-1}$, characteristic for $p$-mode pulsations in $\beta$~Cep stars. The analysis of the photometry separately for NGC\,457-8 and -6661 shows clearly that the variability occurs in NGC\,457-8, the 
brighter star. The difference of 1.57~mag between the two stars leads to a reduction of the amplitudes of variability by about 20 per cent. The parameters of the sine-curve fit to the combined data (with amplitudes corrected for the light dilution) are listed in Table 4. The frequency $f_2$ is subject of the $\pm$1~d$^{-1}$ ambiguity due to strong daily aliases. The $f_1$ agrees within 3$\sigma$ with the value given by Zha12. Their secondary frequency, 4.013~d$^{-1}$, very nearly equal to four times the reciprocal of the sidereal day, is almost certainly spurious. It is not detected in our data down to the detection threshold of 0.7~mmag, well below the amplitude reported by Zha12.

\MakeTable{crcrrrcc}{12.5cm}{Parameters of the sine-curve fits to the 2010\,--\,2011 $B$, $V$ and $I_{\textup{C}}$ differential 
magnitudes of the $\beta$~Cep-type star NGC\,457-8}
{\hline\noalign{\smallskip}
 Filter & $N_{\textup{obs}}$ & \multicolumn{1}{c}{$f$} & \multicolumn{1}{c}{$A$} & \multicolumn{1}{c}{$T_{\textup{max}}$} & $\sigma_{\textup{res}}$\\
  &  & \multicolumn{1}{c}{[d$^{-1}$]} & \multicolumn{1}{c}{[mmag]} & [HJD $-$ 2\,455\,590.0] & [mmag] \\
\noalign{\smallskip}\hline\noalign{\smallskip}
  $B$ &   389 & $f_1$  & 17.55 $\pm$ 0.30 & 11.8770 $\pm$ 0.0002 & 3.32\\
         &	      & $f_2$ & 1.48 $\pm$ 0.30 & 11.9115 $\pm$ 0.0030 &\\
  $V$ & 2048 & $f_1=$ 5.33987 $\pm$ 0.00003& 17.22 $\pm$ 0.12 &  8.8803 $\pm$ 0.0002& 3.10\\
         &	      &$f_2=$ 10.7856 $\pm$ 0.0005 & 1.15 $\pm$ 0.12 &  8.9387 $\pm$ 0.0016 &\\
  $I_{\textup{C}}$ & 404 & $f_1$& 15.73 $\pm$ 0.40 &  8.8801 $\pm$ 0.0008& 4.74\\
         &	      & $f_2$ & 1.26 $\pm$ 0.40 & 8.8565 $\pm$ 0.0049 &\\
\noalign{\smallskip}\hline
}

\subsection{Variable Be Stars}
Be stars are variable on time scales from hours to decades. The erratic variability on time scales of weeks or longer is attributed to the changes in the circumstellar disk and stellar wind. The peak-to-peak amplitudes of these variations usually do not exceed 0.5 mag in the $V$ band. The short-period variability is likely to be caused by pulsations and rotational modulation. Since Be stars occupy the whole range of B spectral types, both $p$-mode ($\beta$~Cep-type) and $g$-mode (SPB-type) pulsations are expected to occur in these stars. 
\begin{figure}[ht]
\begin{center}
\includegraphics[width=\textwidth]{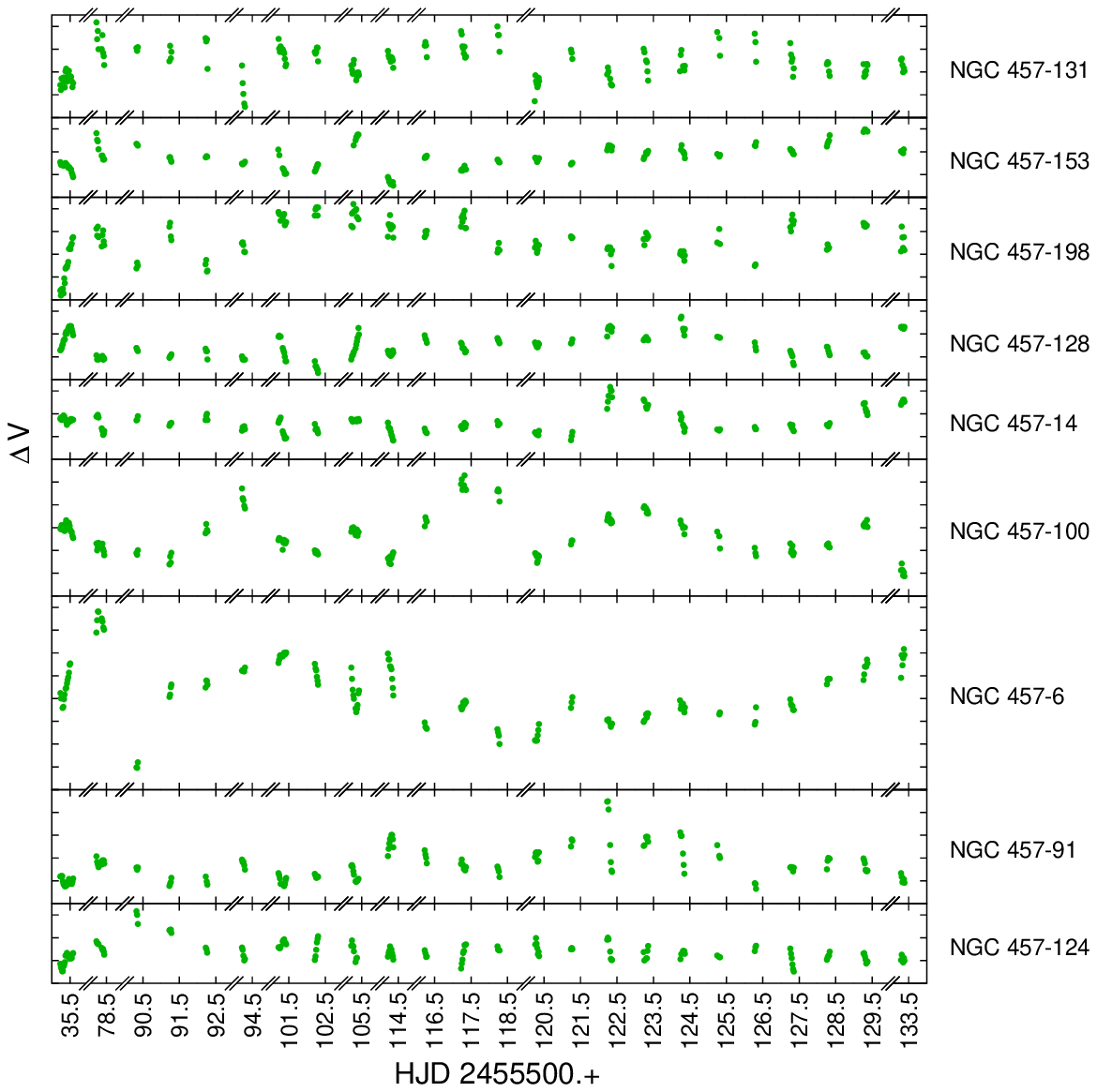}
\end{center}
\FigCap{The $V$-filter light curves of nine variable Be stars in NGC\,457 with the largest amplitudes ($\Delta V>$~0.03~mag). The points were obtained by averaging the original data in 25-min intervals. Only data from the third run (2010\,--\,2011) are shown. Ordinate ticks are separated by 0.02~mag. Note that abscissa is not continuous.}
\end{figure}

In general, variability observed in Be stars appears to be quite complicated. In combination with the severe daily aliasing in our single-site data, the search for periodicities in such stars yields usually uncertain results. Of 15 Be stars listed in Table 1, all but NGC\,457-143 and -87 were found to be variable. The light curves of almost all variable Be stars are dominated by erratic long-term variability. The peak-to-peak amplitudes during the third run do not exceed 0.13~mag in $V$, the shell star NGC\,457-6 showing the largest amplitude (Fig.~9). Light variations of two Be stars, NGC\,457-62 and -43, are dominated by (quasi)periodic variations with frequencies larger than 1~d$^{-1}$. Such Be stars were classified by Walker {\etal}(2005) as SPBe stars and we follow this classification, see Table 3. The three strongest terms in the light curve of NGC\,457-43 have frequencies of 5.3648, 2.5002, and 2.7647~d$^{-1}$. The first frequency is therefore nearly twice higher than the other two. It is interesting 
to note that very often frequencies of sinusoidal terms observed in Be stars form two groups with mean frequency ratio close to two (Uytterhoeven {\etal}2007, Neiner {\etal}2009, Huat {\etal}2009). In observations of the best quality even more groups separated by the same frequency interval occur (Diago {\etal}2009, Balona {\etal}2011). Balona {\etal}(2011) showed that frequency groupings occur also in stars that are not known to be Be stars, but probably are rapid rotators. It is possible that a large group of stars with a similar variability pattern discovered in Magellanic Clouds by Ko{\l}aczkowski {\etal}(2006) are also stars of this type.

In the CM diagram (Fig.~6), the Be stars occupy mainly the upper main sequence of the cluster, though some of them are significantly redder than the main sequence. In particular, NGC\,457-153, the Be star with the strongest emission in H$\alpha$, is much redder than the cluster main sequence (Fig.~6). This is probably a result of circumstellar reddening, a very well known effect observed in many Be stars in open clusters (see, e.g., Fabregat {\etal}1996).

\subsection{SPB Stars}
We classified a star as an SPB star if at least two periodicities with frequencies attributable to $g$ modes were detected above detection threshold in its frequency spectrum and its location in the CM diagram was consistent with a B-type member of the cluster. This resulted in classifying 21 stars as SPB stars (Table 3).  As can be seen in Fig.~6, they occupy the cluster main sequence in the $V$ magnitude range between 11 and 15, corresponding roughly to B3\,--\,B8 range of spectral type. Frequencies of some of SPB stars in Table 5 are quite high. This resembles stars detected by Mowlavi {\etal}(2013) in NGC\,3766. We will discuss these stars in the last section.

Given the severe daily aliasing mentioned earlier, our data are not well suited for searching for periodicities of the order of one day characteristic of SPB stars. If an incorrect frequency is selected as real --- as might have happened in this analysis --- the possibility that frequencies of the next periodic terms are extracted incorrectly increases. For this reason, for periodic stars we do not list all parameters of the (multi)frequency fit. Instead, we provide in Table 5 only frequencies and amplitudes of detected term(s). Due to the aliasing problem, they should be regarded with caution.

In addition to the 21 multiperiodic SPB stars, five other stars can be regarded as SPB candidates. They are classified as periodic (`Per' in Table 3) and lie in the cluster main sequence in the region of B-type stars. These are NGC\,457-122, -93, -2, -126, and -58. 

\MakeOwnTable{rlrl}{12.5cm}{Frequencies and amplitudes of SPB (S), SPBe (e), $\delta$~Sct ($\delta$), $\gamma$~Dor ($\gamma$) and periodic (P) stars in NGC\,457 detected in the $V$-filter data}{\tiny}
{\hline\noalign{\smallskip}
Star & Frequencies [d$^{-1}$] (amplitudes [mmag]) & Star & Frequencies [d$^{-1}$] (amplitudes [mmag]) \\
\noalign{\smallskip}\hline\noalign{\smallskip}
    2 (P) & 3.7593 (1.94)                                                        &   116 (S) & 0.3150 (3.16), 2.2414 (1.67), 4.7173 (1.66), 3.3785 (1.31)\\
    3 (S) & 3.4364 (3.12), 3.3733 (2.16), 4.1243 (1.91),  6.7496 (1.43) &   119 (S) & 1.7848 (2.98), 2.8640 (1.70)\\
    4 (S) & 4.4960 (2.09), 6.9875 (2.01)                                      &   122 (P) & 1.3517 (4.90)\\
   12 (S) & 0.6820 (3.43), 1.4150 (1.38)                                     &   126 (P) & 3.0194 (6.72)\\
   15 (P) & 0.6174 (10.33), 2$f =$ 1.2348 (6.46)                          & 1332 ($\delta$) & 35.4606 (2.98) \\
   17 (S) & 0.3262 (5.42), 1.8861 (3.53), 0.5635 (1.73), 3.6731 (1.50) & 2218 (P) & 1.4544 (5.10)\\
   20 (S) & 2.7396 (5.21), 2.6810 (3.55), 0.8774 (2.49)                   & 2270 (S) & 2.6225 (8.11), 2.3815 (3.96), 0.2978 (2.43)\\
   22 (S) & 1.1255 (2.53), 3.6806 (1.74), 3.8340 (1.35)                   & 2301 ($\gamma$) & 1.4152 (15.29), 1.0307 (8.66) \\
   26 (S) & 4.4694 (3.35), 3.7001 (3.24), 3.0439 (2.02), 6.9342 (1.68) & 2380 ($\delta$) & 35.2518 (3.89)\\
   32 (S) & 1.8531 (3.71), 0.7442 (2.44)                                     & 2424 (P) & 2.4374 (11.93)\\
   34 (S) & 0.5459 (2.18), 0.2610 (1.06)                                     & 3202 (P) & 1.8609 (4.39)\\
   35 (S) & 0.3551 (2.03), 0.7042 (1.66), 1.0812 (1.62), 1.5548 (1.61) & 3233 ($\delta$) & 15.8617 (4.19)\\
   42 (S) & 2.2161 (4.60), 0.8329 (1.61)                                     & 3242 ($\delta$) & 22.1346 (10.67), 20.3684 (6.93), 24.8714 (7.00), \\
   43 (e) & 5.3648 (2.59), 2.7647 (1.41), 2.5004 (1.12)                   &     & 20.6344 (5.92)\\
   48 (S) & 3.3947 (2.54), 3.2917 (1.50), 5.6810 (1.05)                   & 3318 ($\delta$) & 29.4749 (9.98), 2.2510 (2.69)\\
   52 (S) & 4.9018 (4.66), 1.2126 (3.26), 2.6089 (2.41), 5.6835 (1.81),&  3349 (P) & 2.1036 (18.50)\\
           & 2.2420 (1.42)                                                       & 3416 ($\delta$) & 5.2274 (17.45), 20.2622 (8.94), 22.3159 (7.17)\\
   58 (P) & 3.3136 (4.79)                                                       & 3421 (P) & 1.2644 (7.29)\\
   60 (S) & 1.6636 (2.29), 0.1952 (2.01), 4.6740 (1.84)                  & 6166 (S) & 3.7373 (2.54), 2.4894 (2.27)\\
   62 (e) & 1.2407 (2.89), 0.2944 (1.88)                                    & 6373 (P) & 0.2341 (17.60)\\
   63 (P) & 3.5556 (3.69), 2$f =$ 7.1112 (2.26)                           & 6464 (P) & 2.9142 (77.3)\\
   64 (S) & 0.5302 (3.85), 0.7207 (1.64)                                    & 6579 (P) & 3.0516 (35.8)\\
   71 (S) & 0.5679 (2.04), 0.7641 (1.53)                                    & 6697 (P) & 1.3756 (25.22)\\
   84 (S) & 0.8844 (1.65), 0.4973 (1.52)                                     & 7202 ($\delta$) & 10.4932 (8.95), 7.8120 (8.76)\\
   93 (P) & 3.2015 (2.30)  & \\                                                   
\noalign{\smallskip}\hline
}

\subsection{$\delta$ Sct and $\gamma$ Dor Stars}
Seven variable stars in Table 3 are $\delta$ Sct-type pulsating stars. One of them, NGC\,457-3318, has been already discovered by Zha12 (their variable v11). In addition to the mode with frequency of 29.47~d$^{-1}$ they found, we detected a term with frequency of about 2.25~d$^{-1}$. If this periodicity is due to pulsation, the star is a $\delta$ Sct/$\gamma$ Dor hybrid.  Since it lies well outside the $\delta$~Sct instability strip in NGC\,457 (Fig.~6), it is probably a foreground star. Of the remaining six $\delta$~Sct stars, only NGC\,457-2380 is located on the cluster's isochrone. Three other stars, NGC\,457-1332, -3242, and -7202 lie on the cluster double-star sequence, while it is almost certain that the last two $\delta$~Sct stars in our list, NGC\,457-3233 and -3416, are not members of NGC\,457.

The location of NGC\,457-2301 in the CM diagram of NGC\,457 (Fig.~6) and its multiperiodicity (Table 5) prompted us to classify it as $\gamma$ Dor-type variable. Two other monoperiodic stars, NGC\,457-2218 and -2424, are also good candidates to be $\gamma$~Dor stars and members of NGC\,457.

\subsection{Eclipsing Binaries}
In the sample of variable stars we found there are eight eclipsing binaries, including three that were already known. For six, we were able to derive orbital periods; their phased light curves are shown in Fig.\,10. The remaining two are EA-type systems NGC\,457-2330 and -7227. For both these stars only a single eclipse was observed, on HJD\,2455601.2690\,$\pm$\,0.0014 for NGC\,457-2330 and near HJD\,2455627.27 for NGC\,457-7227.

\begin{figure}[ht]
\begin{center}
\includegraphics[width=0.88\textwidth]{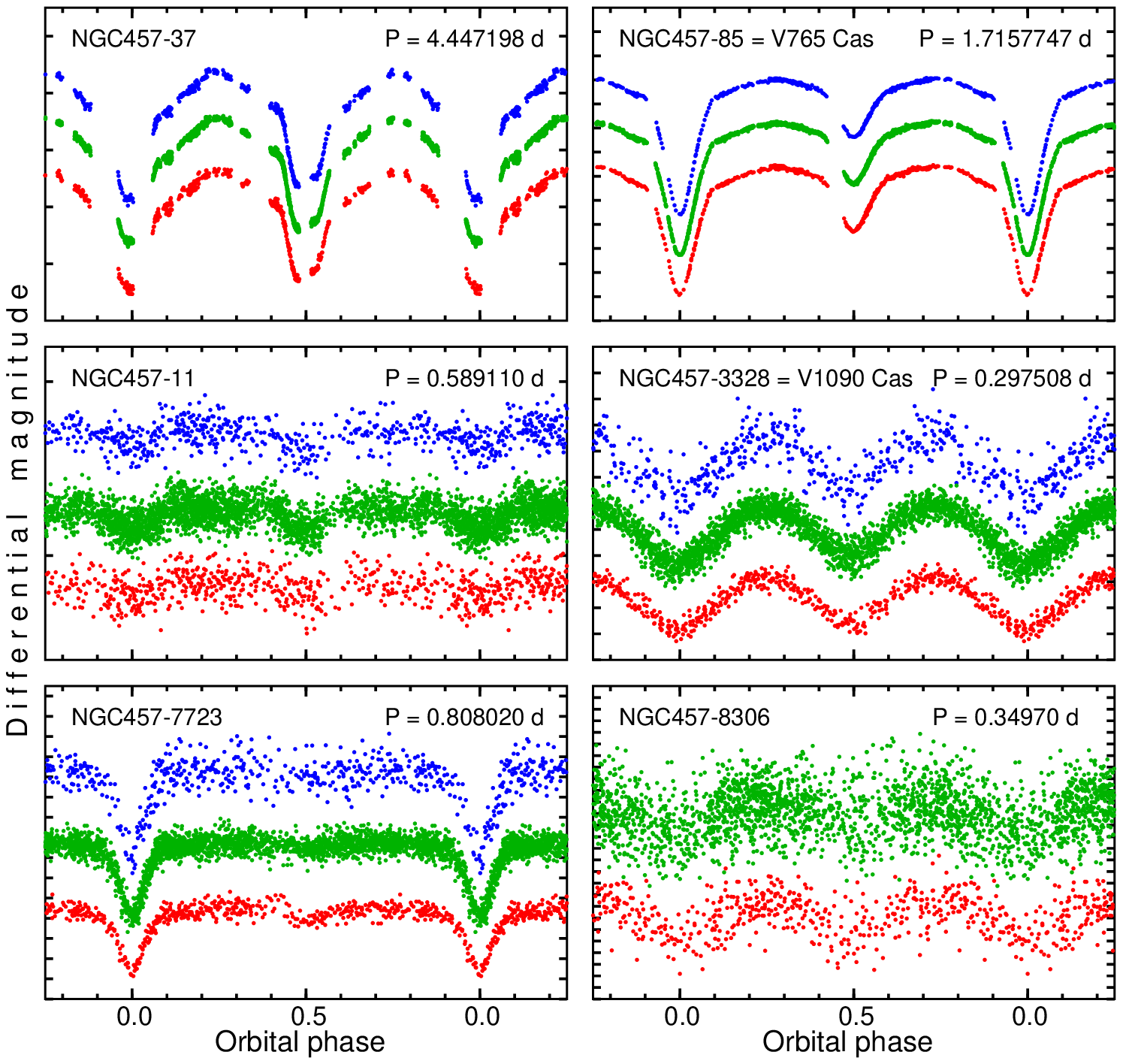}
\end{center}
\FigCap{Phased light curves of six eclipsing binaries in NGC\,457. The blue, green and red dots represent $B$, $V$, and $I_{\rm C}$ data, respectively. The ordinate ticks are separated by 0.1~mag.}
\end{figure}

\MakeTable{rcc}{12.5cm}{Orbital periods and the times of primary minimum for six eclipsing binaries in NGC\,457}
{\hline\noalign{\smallskip}
 & Time of primary minimum & Orbital period \\
Star &  $T_{\rm min}$ I [HJD\,2\,455\,000+] & [d]\\
\noalign{\smallskip}\hline\noalign{\smallskip}
37 & 622.4040 $\pm$ 0.0022 & 4.447198 $\pm$ 0.000009 \\
85 & 601.3317 $\pm$ 0.0005 & 1.7157747 $\pm$ 0.0000028 \\
11 & 535.2700 $\pm$ 0.0012 & 0.589110 $\pm$ 0.000033 \\
3328 & 535.2615 $\pm$ 0.0003 & 0.297508 $\pm$ 0.000009 \\
7723 & 535.3667 $\pm$ 0.0004 & 0.808020 $\pm$ 0.000010 \\
8306 & 535.5497 $\pm$ 0.0020 & 0.34970 $\pm$ 0.00006 \\
\noalign{\smallskip}\hline
}

NGC\,457-37 (BD\,+57$^{\rm o}$249, B1\,V), the brightest eclipsing star in the cluster was discovered by Zha12 (their star v2). It shows a well pronounced reflection effect and a flat-bottom secondary eclipse (better seen in Fig.\,2 of Zha12) which indicates rather small ratio of the radii. Therefore, the small difference of effective temperatures between the components, obtained by Zha12 from light curve modeling, is rather surprising. The system deserves further attention and is the best object in NGC\,457 for which absolute values of masses and radii can be derived. It can be even used for a determination of the distance of the cluster.

The variability of NGC\,457-85 $=$ V765\,Cas $=$ HIP\,6171, the second brightest eclipsing system in NGC\,457, has been discovered by the \textit{Hipparcos} satellite.  With $T_{\rm eff} =$ 22023 $\pm$ 809~K (Fitzpatrick and Massa 2007) the star (or rather its primary) is clearly an early B-type star. The light curve shows proximity effects; the eclipses are partial (Fig.~10).

As can be judged from their position in the cluster CM diagram (Fig.~6), all the remaining six eclipsing binaries but NGC\,457-11 are field objects. NGC\,457-11 (Fig.~10) shows very shallow eclipses and one may doubt if these are eclipses at all. There is also a possibility that the true period is half the value given in Table 6. 

\subsection{The Remaining Variable Stars}
Finally, there are six blue and six red variable stars showing erratic or quasiperiodic variations of light (Table 3). The red variables are located, as expected, redward of the main sequence (Fig.~6), and except for the M-type supergiant NGC\,257-25 (HD\,236697, V466 Cas) were not known as variable prior to our study. V466\,Cas, as mentioned in the Introduction, might be a member of the cluster. The remaining red irregular variables are field stars.

All the irregular blue variables are located in the cluster main sequence. Consequently, they are all possible members. One of these stars, NGC\,457-120 = BD$+$57$^{\rm o}$247, has been classified as an O9.5\,IV star (Hoag and Applequist 1965). It is slightly bluer than the cluster main sequence (see Fig.~6) and therefore was considered to be a blue straggler (Fitzsimmons 1993). However, Huang and Gies (2006a) provide $T_{\rm eff} =$ 22384 $\pm$ 333 K, a value  which is typical rather for a B2 spectral type. Both the proper motion and spectral type of NGC\,457-120 are consistent with membership (Harris 1976). Also its radial velocity (Huang and Gies 2006a) agrees fairly well with the mean radial velocity of the cluster measured to be $-$25~km\,s$^{-1}$ by Liu {\etal}(1989) and $-$33~km\,s$^{-1}$ by Frinchaboy and Majewski (2008). The membership of this star is therefore almost certain.

\section{Discussion and Conclusions}
It turns out that NGC\,457 is very rich in variable stars of spectral type B: out of 79 variable stars we found, 45 are B-type stars. The most interesting is the presence of a group of 21 SPB stars (or even more if candidates are included) which surpasses the population known in other young open clusters like NGC\,7654 (Luo {\etal}2012) or NGC\,6231 (Meingast {\etal}2013). Surprisingly, almost half of them shows frequencies higher than 3~d$^{-1}$. These stars are indicated by an asterisk in the last column of Table 3. Variable SPB stars with relatively high frequencies have also been reported in NGC\,3293 (Handler {\etal}2007, 2008), h~Persei (Majewska-\'Swierzbinowicz {\etal}2008), $\chi$~Persei (Saesen {\etal}2010a, 2013), and recently in NGC\,3766 (Mowlavi {\etal}2013).  It is believed that high frequencies in SPB stars are caused by rapid rotation and can be explained in terms of prograde $g$ modes. A high rotation of B-type stars in NGC\,457 seems to be confirmed by the presence of a large number of Be 
stars as the Be phenomenon is inherently related to fast rotation. Let us also mention the discovery of two SPBe stars (see Sect.~6.2) that may form a link between Be and SPB stars (see Rivinius {\etal}2013). It is therefore reasonable to conclude that the SPB stars with high frequencies are probably fast-rotating SPB stars. A spectroscopic survey is, however, necessary to verify this hypothesis. It is also interesting to note that all SPB stars in NGC\,457 have very low amplitudes: for all but three SPB stars the semi-amplitudes of the modes we detected do not exceed 5~mmag (Table 5). 

As far as the location of variable stars in the CM diagram is concerned, Fig.~6 shows two interesting things. First, there is a clear gap between the lower end of the sequence of SPB stars and the instability strip of $\delta$~Sct stars. This is consistent with the result reported by Balona {\etal}(2011) for the {\it Kepler} B-type stars: they did not find pulsations among stars with the latest subtypes of spectral type B. Second, despite the fact that the detection threshold for stars with $V\sim$16~mag in our data amounts to about 4~mmag, there is only one $\delta$~Sct star, NGC\,457-2380, close to the instability strip that is located on the isochrone. Given the fact that non-pulsating stars are very rare in the $\delta$~Sct instability strip (see, e.g.~Guzik {\etal}2014), we may conclude that the amplitudes of $\delta$~Sct-type pulsations in the members of NGC\,457 are mostly below this detection threshold. Also, the presence of three $\delta$~Sct stars, and a $\gamma$~Dor star and two candidates in the 
double-star sequence contrasts with the lack of variable stars of this type on the isochrone. Going to the upper main sequence, there is only a single $\beta$~Cep-type star in the cluster. This is probably a consequence of both the small population of early B-type stars in NGC\,457 and the fact that the metallicity of the cluster is probably lower than solar (see Sect.~3). 

Summarizing, NGC\,457 is an open cluster rich in both Be and pulsating SPB-type stars. In view of the problems in identification of modes excited in rapidly rotating stars, it will be difficult to perform seismic modeling for SPB stars in NGC\,457. However, an overall picture of stellar variability in this cluster combined with similar surveys in other young clusters, may shed some light on the dependencies between rotation and pulsations in B-type stars. The mode identification in pulsating, rapidly rotating B-type stars is very difficult, but there is some progress. For example, it seems that in some cases it is possible to identify at least the azimuthal orders of the observed modes (Saio 2013) and consequently put a lower limit on the degrees of these modes.

\Acknow{This work was supported by the NCN grant No.~2011/03/ B/ST9/02667 and has received funding from the European Community's Seventh Framework Programme (FP7/2007-2013) under grant agreement no.~269194. Some calculations have been carried out in Wroc{\l}aw Centre for Networking and Supercomputing (http://www.wcss.wroc.pl), grant No.~219. We are indebted to Prof. M.\,Jerzykiewicz for his comments made upon reading the manuscript. We thank Dr.~Jason Jackiewicz for making available telescope time on the APO ARC 3.5-m telescope and for his support during observations. We thank D.\,Drobek, P.\,La\-cho\-wicz, G.\,Michalska, P.\,\'Sr\'odka, and E.\,Zahajkiewicz for making some observations of NGC\,457 in Bia{\l}k\'ow. We also thank Warsaw University Observatory staff for allocating observing time on the 60-cm telescope in Ostrowik. This research has made use of the WEBDA database, operated at the Department of Theoretical Physics and Astrophysics of the Masaryk University.}

\end{document}